\documentclass[12pt]{article}
\usepackage{a4wide}
\usepackage{latexsym}
\usepackage{amsmath}
\usepackage{amsfonts}
\usepackage{amscd}
\usepackage{amsthm}
\usepackage{amssymb}
\usepackage{cite}
\usepackage{shuffle}

\usepackage{pslatex}
\usepackage{graphicx}
\usepackage[latin1,utf8]{inputenc}
\usepackage[T1]{fontenc}

\usepackage{mathtools}
\usepackage{mathdots}

\newcommand{\bq}{\begin{eqnarray}}
\newcommand{\eq}{\end{eqnarray}}
\newcommand{\eps}{\varepsilon}

\newcommand{\qbar}{q}
\newcommand{\Yinvariant}{Y}
\newcommand{\Frobeniusbasis}{\psi}

\theoremstyle{plain}

\usepackage{color}

\allowdisplaybreaks

\begin{document}

\thispagestyle{empty}

\begin{flushright}
  MITP/22-108
 \\ TUM-HEP-1445/22
\end{flushright}

\vspace{1.5cm}

\begin{center}
  {\Large\bf Bananas of equal mass: any loop, any order in the dimensional regularisation parameter\\
  }
  \vspace{1cm}
  {\large Sebastian P\"ogel${}^{a}$, Xing Wang${}^{b}$ and Stefan Weinzierl${}^{a}$\\
  \vspace{1cm}
      {\small \em ${}^{a}$ PRISMA Cluster of Excellence,} \\
      {\small \em Institut f{\"u}r Physik, Staudinger Weg 7,}\\
      {\small \em Johannes Gutenberg-Universit{\"a}t Mainz,}\\
      {\small \em D - 55099 Mainz, Germany}\\
  \vspace{2mm}
      {\small \em ${}^{b}$ ORIGINS Cluster of Excellence,} \\
      {\small \em Physik Department, James-Frank-Stra{\ss}e 1,} \\
      {\small \em Technische Universit\"at M\"unchen,} \\
      {\small \em D - 85748 Garching, Germany}
  } 
\end{center}

\vspace{2cm}

\begin{abstract}\noindent
  {
We describe a systematic approach to cast the differential equation for the $l$-loop equal mass banana integral
into an $\varepsilon$-factorised form.
With the known boundary value at a specific point we obtain systematically the term of order $j$
in the expansion in the dimensional regularisation parameter $\varepsilon$ for any loop $l$.
The approach is based on properties of Calabi--Yau operators, and in particular on self-duality.
   }
\end{abstract}

\vspace*{\fill}

\newpage

\section{Introduction}
\label{sect:intro}

The interplay between physics and geometry is a fascinating topic.
In the context of perturbative quantum field theory
it connects Feynman integrals with the theory of motives and Hodge structures \cite{Bloch:2006}.
Recent advances in our abilities to compute Feynman integrals profited from this geometric insight.
In this paper we push this further:
We present a systematic approach to compute all master integrals of the $l$-loop equal mass banana family
to any order in the dimensional parameter $\eps$.
It is remarkable that this can be done systematically for any loop.
The solution of the master integrals is obtained from an $\eps$-factorised differential equation \cite{Henn:2013pwa}.
We recall that the differential equation in an $\eps$-factorised form 
together with values of the Feynman integrals at a boundary point is all that we need:
From this data we can easily obtain the analytic solution to any order in the dimensional regularisation parameter $\eps$.
This approach has been applied successfully to many Feynman integrals evaluating to multiple polylogarithms
and to several elliptic Feynman integrals \cite{Adams:2018yfj,Bogner:2019lfa,Giroux:2022wav,Muller:2022gec,Dlapa:2022wdu}.
As usual, the bottleneck of any Feynman integral computation is finding a transformation that converts
a non-$\eps$-factorised differential equation into an $\eps$-factorised differential equation.
At this step the input from geometry is extremely helpful: 
For the equal-mass banana integrals we use properties of Calabi--Yau operators to construct this transformation, and here in particular self-duality.

We may associate to any Feynman integral a geometric object and there are many examples
of Feynman integrals whose geometry is given by Calabi--Yau manifolds \cite{Bourjaily:2018yfy}.
In particular, the family of $l$-loop banana integrals provides for $l \ge 2$ 
examples of Feynman integrals that are related to Calabi--Yau $(l-1)$-folds.
This family of integrals has therefore received significant attention 
in recent years \cite{Groote:2005ay,Klemm:2019dbm,Bonisch:2020qmm,Bonisch:2021yfw,Kreimer:2022fxm,Forum:2022lpz}. 
For the $l$-loop banana integrals the geometry is given by an algebraic variety defined by the zero set of
the second graph polynomial in ${\mathbb C}{\mathbb P}^{l}$.

Let us first briefly review the banana integrals at low loop orders.
The one-loop banana integral is rather trivial.
The geometry of the one-loop banana integral --as defined above-- is given by two points, e.g. a zero-dimensional manifold with
two connected components.
Calabi--Yau manifolds are usually assumed to be connected, therefore the geometry of the one-loop banana integral
is not a Calabi--Yau $0$-fold in the strict sense.
It is well-known how to cast the differential equation for the one-loop banana integral into an $\eps$-factorised form,
for a pedagogical discussion see \cite{Weinzierl:2022eaz}.
In this paper we also discuss the one-loop (and zero-loop) banana integral from the perspective of extrapolating
the general all-loop formulae obtained for $l\ge 2$ to the special cases $l=1$ and $l=0$.

The two-loop banana integral is also known as the sunrise integral (or the London transport integral).
It is related to an elliptic curve (a Calabi--Yau $1$-fold).
The sunrise integral has been discussed extensively in the 
literature \cite{Sabry:1962,Broadhurst:1993mw,Laporta:2004rb,Bloch:2013tra,Adams:2015ydq,Adams:2017ejb,Broedel:2017kkb,Broedel:2017siw,Adams:2018yfj,Honemann:2018mrb,Bogner:2019lfa,Giroux:2022wav}.
The $\eps$-factorised form of the differential equation can be found in \cite{Adams:2018yfj}.

The three-loop banana integral is related to a Calabi--Yau $2$-fold.
It has the special property that its Picard--Fuchs operator in two space-time dimensions
is a symmetric square \cite{Verrill:1996,Joyce:1972}.
It can therefore be treated with methods similar to the elliptic case \cite{Bloch:2014qca,Primo:2017ipr,Broedel:2019kmn,Broedel:2021zij,Pogel:2022yat}. 
The $\eps$-factorised form of the differential equation has been given in \cite{Pogel:2022yat}.

The four-loop banana integral has been discussed recently in \cite{Pogel:2022ken},
where also the $\eps$-factorised form of the differential equation has been given.

The available data up to four loops shows that at each new loop order 
there is a new complication not present at previous loop orders.
At one-loop we need a change of variables which rationalises a square root in order to cast the differential
equation into a form which gives harmonic polylogarithms.
At two-loops the transformation of the master integrals is no longer algebraic, but involves transcendental functions, which are
the periods of an elliptic curve.
With an appropriate change of variables the entries of the differential equations are modular forms.
The differential one-forms corresponding to modular forms of modular weight two are all polylogarithmic dlog-forms.
This is a special property at modular weight two, the differential one-forms corresponding to modular forms of modular weight not equal to two are not polylogarithmic dlog-forms.
As the Picard--Fuchs operator at three loops is a symmetric square, the notion of modular weight generalises in a straightforward way to three
loops and we may again look at the entries of modular weight two.
At three loops we see for the first time non-polylogarithmic differential one-forms at modular weight two.
These do not transform as modular forms, but as generalisations thereof. 
In \cite{Pogel:2022yat} they were called ``quasi-Eichler''.
In the notation of this paper it is the statement that the differential one-forms $\tilde{\omega}_{2,j}$ in eq.~(\ref{split_mpl_remainder}) 
may be non-zero for $l \ge 3$.
At four loops we see for the first time so-called $\Yinvariant$-invariants appearing in the $\eps$-factorised differential equation.
We will discuss these in details in section~\ref{sect:structure_series}.

One might guess that this will continue:
that at each new loop order there is a new complication not present at the previous loop order.
The results of this paper show that this is not the case.
The process saturates at four loops and there are no new complications from five loops onwards.
We may therefore give a systematic method to transform the differential equation for the $l$-loop banana integral
into an $\eps$-factorised form.
This method is the main result of this paper.
From the differential equation we may also read off the symbols. 
As a by-product we obtain the symbol alphabet for the $l$-loop equal-mass banana integral,
extending recent work on elliptic symbols \cite{Kristensson:2021ani,Wilhelm:2022wow} to Calabi--Yau manifolds.

With this method and a known boundary value we are able to compute the $l$-loop banana integral.
We do this explicitly for five and six loops.

This paper is organised as follows: 
In section~\ref{sect:defintions} we introduce our notation and the family of the equal mass $l$-loop banana integrals.
In section~\ref{sect:calabi_yau} we discuss Calabi--Yau operators and their self-duality.
Our method for the transformation of the differential equation into an $\eps$-factorised form is given in section~\ref{sect:method}.
In section~\ref{sect:degenerated_cases} we consider the rather simple cases of one and zero loops from the perspective 
of the all-loop order formulae.
The non-trivial examples at two, three and four loops can be found in the literature \cite{Adams:2018yfj,Pogel:2022yat,Pogel:2022ken}.
In section~\ref{sect:5_loops} we treat the equal-mass five-loop banana integral.
In section~\ref{sect:6_loops} we discuss the equal-mass six-loop banana integral. This is the first case involving 
two $\Yinvariant$-invariants $\Yinvariant_2$ and $\Yinvariant_3$.
In section~\ref{sect:relations} we discuss non-trivial relations satisfied by our choice periods.
Finally, our conclusions are given in section~\ref{sect:conclusions}.
In an appendix we review a highly efficient method to derive the differential equation in the derivative basis.
This differential equation is not in an $\eps$-factorised form, but needed as a starting point.


\section{Definitions and conventions}
\label{sect:defintions}

\subsection{The family of banana integrals}

\begin{figure}
\begin{center}
\includegraphics[scale=0.8]{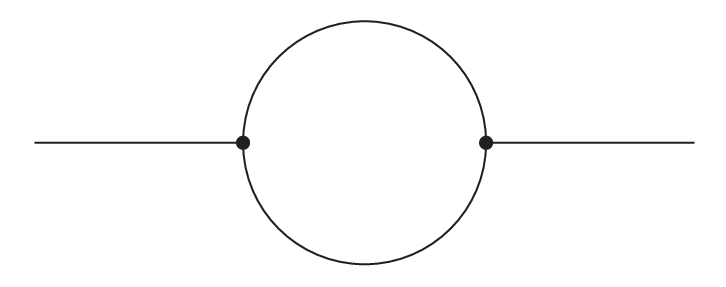}
\includegraphics[scale=0.8]{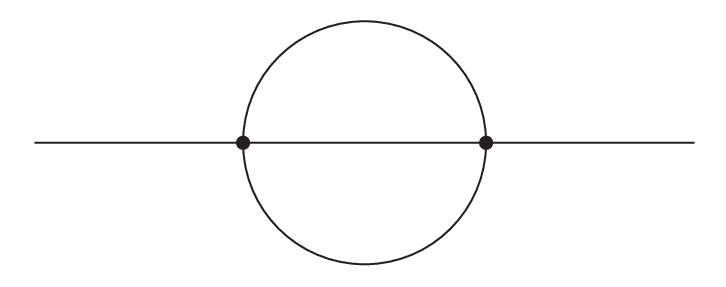}
\\
\vspace*{6mm}
\includegraphics[scale=0.8]{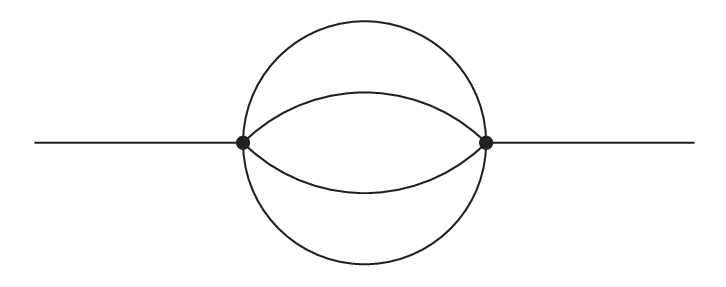}
\includegraphics[scale=0.8]{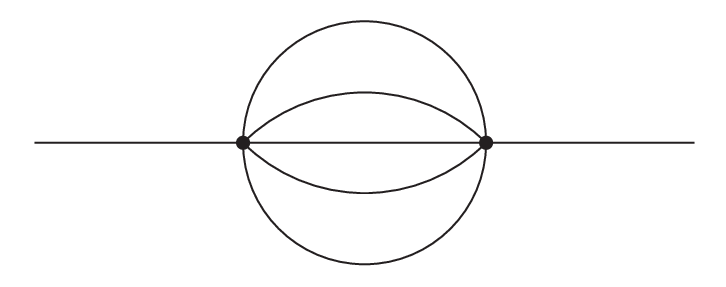}
\end{center}
\caption{
The graphs for the banana integrals from one to four loops.
}
\label{fig_banana_graphs}
\end{figure}

We are interested in the equal mass $l$-loop banana integrals defined by
\bq
\label{def_banana_loop_integral}
 I_{\nu_1 \dots \nu_l \nu_{l+1}}
 & = &
 e^{l \eps \gamma_E}
 \left(m^2\right)^{\nu-\frac{l D}{2}}
 \int \left( \prod\limits_{a=1}^{l+1} \frac{d^Dk_a}{i \pi^{\frac{D}{2}}} \right)
 i \pi^{\frac{D}{2}} \delta^D\left(p-\sum\limits_{b=1}^{l+1} k_b \right)
 \left( \prod\limits_{c=1}^{l+1} \frac{1}{\left(-k_c^2+m^2\right)^{\nu_c}} \right),
\eq
where $D$ denotes the number of space-time dimensions, $\eps$ the dimensional regularisation parameter,
$\gamma_E$ the Euler-Mascheroni constant
and the quantity $\nu$ is defined by
\bq
 \nu & = &
 \sum\limits_{j=1}^{l+1} \nu_j.
\eq
Feynman graphs from one to four loops are shown in fig.~\ref{fig_banana_graphs}.
We consider these integrals in $D=2-2\eps$ space-time dimensions.
As kinematical variables we use 
\bq
 x & = & \frac{p^2}{m^2}
\eq
at finite values of $x$ and
\bq
 y & = & - \frac{1}{x} \;\; = \;\; - \frac{m^2}{p^2}
\eq
around the point $x=\infty$.
It is well known that in the equal mass case there are $(l+1)$ master integrals at $l$ loops.
At $l$ loops a possible basis is given by
\bq
 I_{1 \dots 1 \nu}, & & \nu \in \{0,\dots,l\},
\eq
where $l$ indices $1$ preceed the index $\nu$.
We call this basis the dot basis.
An alternative basis is the derivative basis given by
\bq
\label{def_derivative_basis}
 I_{1 \dots 1 0},
 \;\;\;
 I_{1 \dots 1 1},
 \;\;\;
 \frac{d}{dy}I_{1 \dots 1 1},
 \;\;\;
 \dots,
 \;\;\;
 \frac{d^{l-1}}{dy^{l-1}}I_{1 \dots 1 1}.
\eq
For $l \in \{0,1,2,3\}$ these bases are listed explicitly in table~\ref{table_precanonical_masters}.
\begin{table}
\begin{center}
\begin{tabular}{|l|l|l|}
 \hline
 $l$ & dot basis & derivative basis \\
 \hline
 $0$ & $I_{0}$                               & $I_{0}$ \\ 
 $1$ & $I_{10}, I_{11}$                       & $I_{10}, I_{11}$ \\ 
 $2$ & $I_{110}, I_{111}, I_{112}$             & $I_{110}, I_{111}, \frac{d}{dy} I_{111}$ \\ 
 $3$ & $I_{1110}, I_{1111}, I_{1112}, I_{1113}$ & $I_{1110}, I_{1111}, \frac{d}{dy} I_{1111}, \frac{d^2}{dy^2} I_{1111}$ \\ 
 \hline
\end{tabular}
\end{center}
\caption{
Possible bases of master integrals for $l \in \{0,1,2,3\}$.
}
\label{table_precanonical_masters}
\end{table}
We may include the trivial $0$-loop case.
Note that at $0$ loops eq.~(\ref{def_banana_loop_integral}) gives
\bq
 I_\nu
 & = &
 \left( \frac{1}{1-x} \right)^\nu
\eq
and in particular
\bq
 I_0
 & = &
 1.
\eq
We denote by $M^{(l)}=(M^{(l)}_{0}, M^{(l)}_{1}, \dots, M^{(l)}_{l})^T$ a basis of master integrals at $l$ loops,
such that the differential equation is in $\eps$-factorised form.
The main result of this paper is a systematic procedure to construct this basis.
The $k$-th master integral at $l$ loops is denoted by
\bq
 M^{(l)}_{k}
\eq
and its $\eps$-expansion by
\bq
 M^{(l)}_{k}
 & = & 
 \sum\limits_{j=0}^\infty M^{(l,j)}_{k} \eps^j.
\eq
If it is clear from the context that we are considering a fixed loop order $l$ we drop the superscript $(l)$
to simplify the notation
and write for example 
\bq
 M & = & \left(M_{0}, M_{1}, \dots, M_{l}\right)^T
\eq
for a basis at $l$ loops.

\subsection{Calabi--Yau geometry}

The $l$-loop banana integral is related to a Calabi--Yau $(l-1)$-fold for $l \ge 2$.
This is most easily seen in the Feynman parameter representation, which is given for
the $l$-loop banana integral by
\bq
\label{banana_Feynman_parametrisation}
 I_{\nu_1 \dots \nu_l \nu_{l+1}}
 & = &
 \frac{e^{l \eps \gamma_E}
 \Gamma\left(\nu-\frac{l D}{2}\right)}{\prod\limits_{j=1}^{l+1}\Gamma(\nu_j)}
 \int\limits_{\Delta} \omega \;
 \left( \prod\limits_{j=1}^{l+1} a_j^{\nu_j-1} \right)
 \frac{{\mathcal U}^{\nu-\frac{\left(l+1\right) D}{2}}}{{\mathcal F}^{\nu-\frac{l D}{2}}},
\eq
with $\Delta={\mathbb R} {\mathbb P}^{l}_{\ge 0}$ and
\bq
 \omega & = & \sum\limits_{j=1}^{l+1} (-1)^{j-1}
  \; a_j \; da_1 \wedge ... \wedge \widehat{da_j} \wedge ... \wedge da_n.
\eq
The hat indicates that the corresponding term is omitted.
The graph polynomials are given by
\bq
\label{def_graph_polynomials}
 {\mathcal U}
 \; = \;
 \left( \prod\limits_{i=1}^{l+1} a_i \right) \cdot \left( \sum\limits_{j=1}^{l+1} \frac{1}{a_j} \right),
 & &
 {\mathcal F}
 \; = \;
 - x \left( \prod\limits_{i=1}^{l+1} a_i \right)
 + \left( \sum\limits_{i=1}^{l+1} a_i \right) {\mathcal U}.
\eq
At one, two and three loops we have for the second graph polynomial
\bq
 l=1 & : & 
 {\mathcal F} = - a_1 a_2 x + \left( a_1 + a_2 \right)^2,
 \\
 l=2 & : & 
 {\mathcal F} = - a_1 a_2 a_3 x + \left( a_1 a_2 + a_1 a_3 + a_2 a_3 \right) \left( a_1 + a_2 + a_3 \right),
 \nonumber \\
 l=3 & : & 
 {\mathcal F} = - a_1 a_2 a_3 a_4 x + \left( a_1 a_2 a_3 + a_1 a_2 a_4 + a_1 a_3 a_4 + a_2 a_3 a_4 \right) \left( a_1 + a_2 + a_3 + a_4 \right).
 \nonumber
\eq
For $D=2$ space-time dimensions eq.~(\ref{banana_Feynman_parametrisation}) reduces to
\bq
 I_{\nu_1 \dots \nu_l \nu_{l+1}}
 & = &
 \int\limits_{\Delta} \frac{\omega}{{\mathcal F}}.
\eq
The geometry of the banana integrals is determined by the variety where ${\mathcal F}$ vanishes:
\bq
 X & = & 
 \left\{ \; \left[a_1:a_2:\dots:a_{l+1}\right] \, \in \, {\mathbb C} {\mathbb P}^{l} \; | \; {\mathcal F}(a) \, = \, 0 \; \right\}.
\eq
The second graph polynomial is a homogeneous polynomial of degree $(l+1)$.
For generic values of the variable $x$ the hypersurface $X \in {\mathbb C} {\mathbb P}^{l}$ 
is smooth for $l \in \{1,2\}$ and singular for $l \ge 3$:
At $l$ loops (with $l \ge 2$) and for generic $x$ the hypersurface $X$ has points of multiplicity $(l-1)$.
For example, at three loops and for generic $x$ the point $[1:0:0:0]$ is a point of multiplicity $2$, as ${\mathcal F}$ 
and the first derivatives $\partial {\mathcal F}/\partial a_i$ vanish there (and there is a non-vanishing second derivative).
It is possible to relate $X$ by a birational map to a smooth variety, see for example \cite{Klemm:2019dbm,Bonisch:2020qmm,Bonisch:2021yfw,Candelas:2021lkc}.
This allows us to conclude that $X$ defines for $l \ge 2$ a Calabi--Yau $(l-1)$-fold.
In particular we have at two loops an elliptic curve and at three loops a K3 surface.

\subsection{Singularities}

We study the family of banana integrals through their differential equations.
The differential equation will have regular singular points.
For example, the differential equation for the two-loop banana integral (the sunrise integral)
has in $x$-space regular singular points at
\bq
 \left\{ 0, 1, 9, \infty \right\}.
\eq
The regular singular point $x=9$ corresponds to the threshold $p^2=(m+m+m)^2$,
the regular singular point $x=1$ corresponds to the pseudo-threshold $p^2=(m+m-m)^2$.
It is not too difficult to derive the set of all possible singularities of the differential equation.
Apart from the points $0$ and $\infty$ they are given by the threshold and the pseudo-thresholds,
which can be obtained by considering all sign choices of
\bq
 p^2 & = & \left( m \pm m \pm \dots \pm m \right)^2,
\eq
with $(l+1)$ summands inside the bracket on the right-hand side.

We denote by $S^{(l)}$ the set of singular points not equal to $0$ nor $\infty$ of the differential equation at $l$ loops in the $x$-coordinate system.
We have to distinguish the cases where $l$ is odd or even.
For $l \ge 1$ we have
\bq
 S^{(l)}
 & = &
 \left\{ \begin{array}{ll}
  \left\{ \left(2k\right)^2 | k \in \left\{1,\dots,\frac{l+1}{2} \right\} \right\}, & l \; \mbox{odd}, \\
  \left\{ \left(2k-1\right)^2 | k \in \left\{1,\dots,\frac{l+2}{2} \right\} \right\}, & l \; \mbox{even}. \\
 \end{array} \right.
\eq
We may extend the definition to $l=0$.
\begin{table}
\begin{center}
\begin{tabular}{|l|l|}
 \hline
 $l$ & $S^{(l)}$ \\
 \hline
 $0$ & $\left\{ 1 \right\}$ \\ 
 $1$ & $\left\{ 4 \right\}$ \\ 
 $2$ & $\left\{ 1, 9 \right\}$ \\ 
 $3$ & $\left\{ 4, 16 \right\}$ \\ 
 $4$ & $\left\{ 1, 9, 25 \right\}$ \\ 
 $5$ & $\left\{ 4, 16, 36 \right\}$ \\ 
 \hline
\end{tabular}
\end{center}
\caption{
The set $S^{(l)}$ for $l \in \{0,1,2,3,4,5\}$.
}
\label{table_singular_points}
\end{table}
For $l \le 5$ the sets $S^{(l)}$ are listed in table~\ref{table_singular_points}.

\subsection{Picard--Fuchs operators}
\label{sect:picard_fuchs_operator}

At $l$ loops we consider the integral
\bq
 I_{1\dots11},
\eq
where all propagators occur to the power one.
This integral satisfies a linear inhomogeneous differential equation of order $l$:
\bq
\label{full_Picard_Fuchs_dgl}
 L^{(l)} I_{1\dots11}
 & = &
 \left(-1\right)^l \frac{\left(l+1\right)! }{y^{l-1} \prod\limits_{a \in S^{(l)}} \left(1+ay\right)}
 \eps^l I_{1\dots10},
\eq
with
\bq
 L^{(l)} 
 & = &
 \frac{d^l}{dy^l} + \sum\limits_{j=0}^{l-1} r^{(l)}_j \frac{d^j}{dy^j}.
\eq
The differential operator $L^{(l)}$ is called the Picard--Fuchs operator for the integral $I_{1\dots11}$.
An efficient method to compute the Picard--Fuchs operator is reviewed in appendix~\ref{sect:bessel}.
The coefficients $r_j$ are rational functions in $y$ and polynomials in $\eps$.
We denote by $L^{(l,0)}$ the $\eps^0$-part of $L^{(l)}$.
We write
\bq
\label{def_Picard_Fuchs_eps_0}
 L^{(l,0)} 
 & = &
 \frac{d^l}{dy^l} + \sum\limits_{j=0}^{l-1} r^{(l,0)}_j \frac{d^j}{dy^j}.
\eq
The operator $L^{(l,0)}$ plays an important role in constructing a basis, which leads to an $\eps$-factorised differential
equation.
Up to four loops we have
\bq
\label{examples_Picard_Fuchs}
 L^{(1,0)} 
 & = &
 \frac{d}{dy} - \frac{1}{y} + \frac{1}{2} \frac{4}{\left(1+4y\right)},
 \\
 L^{(2,0)} 
 & = &
 \frac{d^2}{dy^2} + \left[ - \frac{1}{y} + \frac{1}{1+y} + \frac{9}{1+9y} \right] \frac{d}{dy}
 + \frac{1+3y}{y^2\left(1+y\right)\left(1+9y\right)},
 \nonumber \\
 L^{(3,0)} 
 & = &
 \frac{d^3}{dy^3} + \left[ \frac{3}{2} \frac{4}{\left(1+4y\right)} + \frac{3}{2} \frac{16}{\left(1+16y\right)} \right] \frac{d^2}{dy^2}
 + \frac{1+8y+64y^2}{y^2\left(1+4y\right)\left(1+16y\right)} \frac{d}{dy}
 - \frac{1}{y^3\left(1+16y\right)}.
 \nonumber \\
 L^{(4,0)} 
 & = &
 \frac{d^4}{dy^4} + \left[ \frac{2}{y} + 2 \frac{1}{\left(1+y\right)} + 2 \frac{9}{\left(1+9y\right)}  + 2 \frac{25}{\left(1+25y\right)} \right] \frac{d^3}{dy^3}
 \nonumber \\
 & &
 + \frac{1+98y+1839y^2+3150y^3}{y^2\left(1+y\right)\left(1+9y\right)\left(1+25y\right)} \frac{d^2}{dy^2}
 - \frac{\left(1-15y-60y^2\right)\left(1+15y\right)}{y^3\left(1+y\right)\left(1+9y\right)\left(1+25y\right)} \frac{d}{dy}
 \nonumber \\
 & &
 + \frac{1+5y}{y^4\left(1+y\right)\left(1+9y\right)\left(1+25y\right)}.
\eq
The general form of the coefficient of the second-to-highest derivative is
\bq
 r^{(l,0)}_{l-1}
 & = &
 \frac{l \left(l-3\right)}{2 y}
 + \frac{l}{2} \sum\limits_{a \in S^{(l)}} \frac{a}{1+ay}.
\eq
We then consider the differential equation 
\bq
\label{homogeneous_diff_eq}
 L^{(l,0)} \Frobeniusbasis^{(l)} & = & 0.
\eq
This is a homogeneous linear differential equation of order $l$.
In the language of physics it is the differential equation satisfied by the maximal cut of the $l$-loop banana integral
in $D=2$ space-time dimensions.
We denote the $l$ independent solutions by $\Frobeniusbasis^{(l)}_0, \Frobeniusbasis^{(l)}_1, \dots, \Frobeniusbasis^{(l)}_{l-1}$.

The indicial equation for the operator $L^{(l,0)}$ at the point $y=0$ is $(\rho-1)^l=0$, 
showing that $y=0$ is a point of maximal unipotent monodromy.
From the method of Frobenius it follows that we may write the 
$l$ independent solutions $\Frobeniusbasis^{(l)}_0-\Frobeniusbasis^{(l)}_{l-1}$ as
\bq
\label{Frobenius_series}
 \Frobeniusbasis^{(l)}_k 
 & = &
 \frac{1}{\left(2\pi i\right)^{k}}
 \sum\limits_{j=0}^{k}
 \frac{\ln^j y}{j!} 
 \sum\limits_{n=0}^\infty 
 a^{(l)}_{k-j,n} y^{n+1}.
\eq
As normalisation we choose $a^{(l)}_{0,0}=1$.
The solution $\Frobeniusbasis^{(l)}_0$ is holomophic at $y=0$ and we call this solution 
the holomorphic solution.
The holomorphic solution $\Frobeniusbasis^{(l)}_0$ is given for $l \ge 1$ by
\bq
\label{Frobenius_series_holomorphic}
 \Frobeniusbasis^{(l)}_0
 & = &
 \sum\limits_{n=0}^\infty 
 a^{(l)}_{0,n} y^{n+1},
\eq
with
\bq
\label{def_a_l_0_n}
 a^{(l)}_{0,n}
 & = &
 \left(-1\right)^n
 \sum\limits_{n_1+\ldots+n_{l+1}=n}
 \left( \frac{n!}{n_1! \cdots n_{l+1}!} \right)^2.
\eq
Explicitly, we have for the first few terms at low loop orders:
\bq
\label{examples_Frobenius_0}
 \Frobeniusbasis_0^{(1)}
 & = &
 y \left( 1 - 2 y + 6 y^2 - 20 y^3 + 70 y^4 - 252 y^5 \right)
 + {\mathcal O}\left(y^7\right),
 \nonumber \\
 \Frobeniusbasis_0^{(2)}
 & = &
 y \left( 1 - 3 y + 15 y^2 - 93 y^3 + 639 y^4 - 4653 y^5 \right)
 + {\mathcal O}\left(y^7\right),
 \nonumber \\
 \Frobeniusbasis_0^{(3)}
 & = &
 y \left( 1 - 4 y + 28 y^2 - 256 y^3 + 2716 y^4 - 31504 y^5 \right)
 + {\mathcal O}\left(y^7\right),
 \nonumber \\
 \Frobeniusbasis_0^{(4)}
 & = &
 y \left( 1 - 5 y + 45 y^2 - 545 y^3 + 7885 y^4 - 127905 y^5 \right)
 + {\mathcal O}\left(y^7\right),
 \nonumber \\
 \Frobeniusbasis_0^{(5)}
 & = &
 y \left( 1 - 6 y + 66 y^2 - 996 y^3 + 18306 y^4 - 384156 y^5 \right)
 + {\mathcal O}\left(y^7\right),
 \nonumber \\
 \Frobeniusbasis_0^{(6)}
 & = &
 y \left( 1 - 7 y + 91 y^2 - 1645 y^3 + 36715 y^4 - 948157 y^5 \right)
 + {\mathcal O}\left(y^7\right).
\eq
For $l\ge 2$ we have at least two solutions and we call $\Frobeniusbasis^{(l)}_1$ the single-logarithmic solution.
There is also an all-loop formula for the single-logarithmic solution $\Frobeniusbasis^{(l)}_1$.
We write
\bq
\label{Frobenius_series_single_log}
 \Frobeniusbasis^{(l)}_1
 & = &
 \frac{1}{\left(2\pi i\right)}
 \sum\limits_{n=0}^\infty 
 \left[
 a^{(l)}_{1,n} 
 +
 a^{(l)}_{0,n} 
 \ln y
 \right]
 y^{n+1}.
\eq
The coefficients $a^{(l)}_{0,n}$ are the ones given in eq.~(\ref{def_a_l_0_n}).
The coefficients $a^{(l)}_{1,n}$ are given by
\bq
\label{def_a_l_1_n}
 a^{(l)}_{1,n}
 & = &
 2 \left(-1\right)^n
 \sum\limits_{n_1+\ldots+n_{l+1}=n}
 \left( \frac{n!}{n_1! \cdots n_{l+1}!} \right)^2
 \left[ S_1\left(n\right) - S_1\left(n_1\right) \right],
\eq
where $S_1(n)$ denotes the harmonic sum
\bq
 S_1\left(n\right) & = &  \sum\limits_{j=1}^n \frac{1}{j}.
\eq
The holomorphic solution $\Frobeniusbasis^{(l)}_0$ and the single-logarithmic solution $\Frobeniusbasis^{(l)}_1$
are used to define a change of variables from $y$ to $\tau^{(l)}$ (or $\qbar^{(l)}$).
We set
\bq
\label{mirror_map}
 \tau^{(l)} = \frac{\Frobeniusbasis^{(l)}_1}{\Frobeniusbasis^{(l)}_0},
 & &
 \qbar^{(l)} = e^{2\pi i \tau^{(l)}}.
\eq
In the context of Calabi--Yau manifolds the map from $y$ to $\tau^{(l)}$ is called the mirror map \cite{Candelas:1990rm,Batyrev:1993oya,Batyrev:1995ca}.
In the special case of $l=2$ the map corresponds to the transformation from $y$ to the modular
parameter $\tau^{(2)}$ of an elliptic curve.
We denote the Jacobian of the transformation in eq.~(\ref{mirror_map}) by
\bq
\label{def_Jacobian}
 J^{(l)} & = & \frac{1}{2\pi i} \frac{dy}{d\tau^{(l)}},
\eq
the additional factor of $(2\pi i)$ is a convenient convention as it eliminates factors of $(2\pi i)$  in 
subsequent formulae.
From the definition we have
\bq
 J^{(l)}
 & = & 
 \frac{1}{2 \pi i} \frac{\left(\Frobeniusbasis^{(l)}_0\right)^2}{\left(\Frobeniusbasis^{(l)}_0 \partial_y \Frobeniusbasis^{(l)}_1 - \Frobeniusbasis^{(l)}_1 \partial_y \Frobeniusbasis^{(l)}_0\right)}.
\eq
The map from $y$ to $\qbar^{(l)}$ can be inverted, yielding $y$ as a power series in $\qbar^{(l)}$.
Although the differential equation in eq.~(\ref{homogeneous_diff_eq}) has only for $l \ge 2$ a solution space
of dimensions two or greater, we will discuss in detail in section~\ref{sect:degenerated_cases} 
that we may extend the change of variables to $l=1$ and $l=0$.
Doing so and expressing $y$ as a power series in $\qbar^{(l)}$ we find up to six loops
(for better readability we simply write $\qbar$ instead of $\qbar^{(l)}$)
\bq
\label{examples_y_qbar_expansion}
 l=0:
 & &
 y
 =
 \qbar,
 \\
 l=1:
 & & 
 y
 = 
 \qbar + 2 \qbar^2 + 3 \qbar^3 + 4 \qbar^4 + 5 \qbar^5 + 6 \qbar^6 
 + {\mathcal O}\left(\qbar^7\right),
 \nonumber \\
 l=2:
 & & 
 y
 = 
 \qbar + 4 \qbar^2 + 10 \qbar^3 + 20 \qbar^4 + 39 \qbar^5 + 76 \qbar^6 
 + {\mathcal O}\left(\qbar^7\right),
 \nonumber \\
 l=3:
 & &
 y
 = 
 \qbar + 6 \qbar^2 + 21 \qbar^3 + 68 \qbar^4 + 198 \qbar^5 + 510 \qbar^6 
 + {\mathcal O}\left(\qbar^7\right),
 \nonumber \\
 l=4:
 & &
 y
 = 
 \qbar + 8 \qbar^2 + 36 \qbar^3 + 168 \qbar^4 + 514 \qbar^5 + 2760 \qbar^6 
 + {\mathcal O}\left(\qbar^7\right),
 \nonumber \\
 l=5:
 & &
 y
 = 
 \qbar + 10 \qbar^2 + 55 \qbar^3 + 340 \qbar^4 + 955 \qbar^5 + 13222 \qbar^6 
 + {\mathcal O}\left(\qbar^7\right).
 \nonumber \\
 l=6:
 & &
 y
 = 
 \qbar + 12 \qbar^2 + 78 \qbar^3 + 604 \qbar^4 + 1425 \qbar^5 + 47028 \qbar^6 
 + {\mathcal O}\left(\qbar^7\right).
 \nonumber
\eq


\section{Calabi--Yau operators and duality}
\label{sect:calabi_yau}

The Picard--Fuchs operator $L^{(l,0)}$ of eq.~(\ref{def_Picard_Fuchs_eps_0}) is a Calabi--Yau operator.
In this section we review the definition of Calabi--Yau operators and their main properties.
This section is based on \cite{2013arXiv1304.5434B},
more mathematical literature can be found in refs.~\cite{Morrison:1991cd,Ceresole:1992su,Almkvist:2004kj,Almkvist:2005,Yang:2008,2011arXiv1105.1136B,2017arXiv170400164V,Candelas:2021tqt}.

In this section we use the following notation: If $L$ is a differential operator in the variable $y$, and $f(y)$
a function of $y$, then $L(f(y))$ denotes the function obtained by applying $L$ to $f$.
On the other hand, $L f(y)$ (e.g. without brackets) or simply $L f$ 
denotes the differential operator obtained by multiplying $L$
with $f$ from the right.
 
\subsection{Essentially self-adjoint operators}

We consider the differential operator
\bq
\label{generic_differential_operator}
 L
 & = &
 \sum\limits_{j=0}^{l} r_j\left(y\right) \frac{d^j}{dy^j}.
\eq
The adjoint operator $L^\ast$ of the operator $L$ is defined to be
\bq
 L^\ast
 & = &
 \sum\limits_{j=0}^{l} \left(-1\right)^{l-j} \frac{d^j}{dy^j} r_j\left(y\right),
\eq
where the derivatives now also act on the coefficients $r_j(y)$.
An operator $L$ is called self-adjoint, if $L^\ast=L$.
An operator $L$ is called essentially self-adjoint, if there
exists a function $\alpha(y)$ such that
\bq
\label{def_self_adjoint}
 \alpha L^\ast
 & = & 
 L \alpha.
\eq 
An essentially self-adjoint operator is also called a self-dual operator.
If an operator is essentially self-adjoint, the corresponding
$\alpha(y)$ is the solution of the differential equation
\bq
 \frac{d}{dy} \alpha
 & = & 
 \left( -\frac{2}{l}\frac{r_{l-1}}{r_{l}} + \frac{r_l'}{r_l} \right)\alpha,
\eq
where $r_l'=\frac{d}{dy}r_l$.
This differential equation can easily be obtained by comparing the
coefficients of $\frac{d^{l-1}}{dy^{l-1}}$
of both sides of eq.~\ref{def_self_adjoint}.

The Picard--Fuchs operator $L^{(l,0)}$ of eq.~(\ref{def_Picard_Fuchs_eps_0}) is essentially self-adjoint
with
\bq
\label{def_alpha_banana}
 \alpha
 & = &
 \frac{1}{y^{l-3} \prod\limits_{a \in S^{(l)}} \left(1+ay\right)}.
\eq
Note that the $\eps$-dependent Picard--Fuchs operators $L^{(l)}$ are in general not essentially self-adjoint.
Although $L^{(1)}$ and $L^{(2)}$ are essentially self-adjoint, this is no longer true for $l \ge 3$ \cite{Bonisch:2021yfw}.

\subsection{The structure series}
\label{sect:structure_series}

Let $\theta =y \frac{d}{dy}$ denote the Euler operator.
Consider a differential operator $L$ as in eq.~(\ref{generic_differential_operator}) of order $l$ and assume that 
$L$ is self-dual and that
$y=0$ 
is a point of maximal unipotent monodromy.
Let $\Frobeniusbasis_0, \dots, \Frobeniusbasis_{l-1}$ be a Frobenius basis.

We define recursively operators $N_j$ by
\bq
 N_0 \; = \; 1,
 & &
 N_{j+1} \; = \; \theta \frac{1}{\left(2\pi i \right)^j N_j\left(\Frobeniusbasis_{j}\right)} N_j.
\eq
We further set
\bq
 \alpha_j
 & = &
 \frac{1}{\left( 2\pi i \right)^j}
 \frac{1}{N_j\left(\Frobeniusbasis_{j}\right)}.
\eq
With this definition we have for $j \ge 0$
\bq
 N_{j+1} & = & \theta \alpha_j N_j.
\eq
In this way we obtain differential operators $N_0,N_1,\dots,N_l$.
The operators $N_j$ have the property that
\bq
 N_j\left(\Frobeniusbasis_{i}\right)
 & = & 0
 \;\;\;
 \mbox{for} \;\; i \; < \; j.
\eq
We call the sequence $(\alpha_1,\alpha_2,\dots,\alpha_{l-1})$ the structure series of the differential
operator $L$.

As an example we consider the Picard--Fuchs operators $L^{(l,0)}$ of eq.~(\ref{def_Picard_Fuchs_eps_0}).
Up to $4$ loops we find
\bq
\label{examples_structure_series}
 l \; = \; 2: 
 & & 
 \alpha_1^{(2)} 
 \; = \; 
 1 + 4 y - 12 y^2 + 60 y^3 - 348 y^5 + 2196 y^5
 + {\mathcal O}\left(y^6\right),
 \nonumber \\
 l \; = \; 3: 
 & & 
 \alpha_1^{(3)} 
 \; = \; 
 1 + 6 y - 30 y^2 + 276 y^3 - 3030 y^5 + 36012 y^5
 + {\mathcal O}\left(y^6\right),
 \nonumber \\
 & &
 \alpha_2^{(3)} \; = \;\alpha_1^{(3)},
 \nonumber \\
 l \; = \; 4: 
 & & 
 \alpha_1^{(4)} 
 \; = \; 
 1 + 8 y - 56 y^2 + 760 y^3 - 12760 y^5 + 236488 y^5
 + {\mathcal O}\left(y^6\right),
 \nonumber \\
 & & 
 \alpha_2^{(4)} 
 \; = \; 
 1 + 9 y - 72 y^2 + 1080 y^3 - 19248 y^5 + 369936 y^5 
 + {\mathcal O}\left(y^6\right),
 \nonumber \\
 & &
 \alpha_3^{(4)} \; = \;\alpha_1^{(4)}.
\eq
We further define for $j \in \{1,\dots,l-1\}$
\bq
 \Yinvariant_j
 & = & \frac{\alpha_1}{\alpha_j}.
\eq
The function $\Yinvariant_j$ is called the 
$j$-th $\Yinvariant$-invariant\footnote{In ref.~\cite{2013arXiv1304.5434B} the $\Yinvariant$-invariants are denoted with a shift in the index:
Our $\Yinvariant_j$ is denoted as $\Yinvariant_{j-1}$ there.} 
of $L$.

For the structure series we have the symmetry
\bq
 \alpha_j & = & \alpha_{l-j},
\eq
this translates to the symmetry 
\bq
 \Yinvariant_j & = & \Yinvariant_{l-j}
\eq
for the $\Yinvariant$-invariants.

Working out the first few cases we find (with $\tau$ defined by eq.~(\ref{mirror_map}))
\bq
 \Yinvariant_1 = 1,
 \;\;\;\;
 \Yinvariant_2 = \frac{d^2}{d\tau^2} \frac{\Frobeniusbasis_2}{\Frobeniusbasis_0},
 \;\;\;\;
 \Yinvariant_3 = \frac{d}{d\tau}\left( \frac{1}{\Yinvariant_2} \frac{d^2}{d\tau^2} \frac{\Frobeniusbasis_3}{\Frobeniusbasis_0} \right),
 \;\;\;\;
 \Yinvariant_4 = \frac{d}{d\tau}\left[ \frac{1}{\Yinvariant_3} \frac{d}{d\tau}\left( \frac{1}{\Yinvariant_2} \frac{d^2}{d\tau^2} \frac{\Frobeniusbasis_4}{\Frobeniusbasis_0} \right)\right].
\eq
The higher $\Yinvariant$-invariants can be worked out analogously.

As an example we consider again the Picard--Fuchs operators $L^{(l,0)}$ of eq.~(\ref{def_Picard_Fuchs_eps_0}).
At four and five loops we need the non-trivial $\Yinvariant$-invariant $\Yinvariant_2$,
at six and seven loops we need the non-trivial $\Yinvariant$-invariants $\Yinvariant_2$ and $\Yinvariant_3$.
The number of required $\Yinvariant$-invariants increases by one whenever we increase the loop number from odd to even.
The first non-trivial examples are
(again we write for better readability simply $\qbar$ instead of $\qbar^{(l)}$)
\bq
 l \; = \; 4: 
 & & 
 \Yinvariant_2^{(4)} 
 \; = \; 
 1 - \qbar + 17 \qbar^2 - 253 \qbar^3 + 3345 \qbar^4 - 43751 \qbar^5 
 + {\mathcal O}\left(\qbar^6\right),
 \nonumber \\
 l \; = \; 5: 
 & & 
 \Yinvariant_2^{(5)} 
 \; = \; 
 1 - 2 \qbar + 46 \qbar^2 - 1010 \qbar^3 + 21550 \qbar^4 - 463502 \qbar^5 
 + {\mathcal O}\left(\qbar^6\right),
 \nonumber \\
 l \; = \; 6: 
 & & 
 \Yinvariant_2^{(6)} 
 \; = \; 
 1 - 3 \qbar + 87 \qbar^2 - 2523 \qbar^3 + 74247 \qbar^4 - 2248278 \qbar^5 
 + {\mathcal O}\left(\qbar^6\right),
 \nonumber \\
 & & 
 \Yinvariant_3^{(6)} 
 \; = \; 
 1 - 4 \qbar + 124 \qbar^2 - 3892 \qbar^3 + 123564 \qbar^4 - 3985904 \qbar^5 
 + {\mathcal O}\left(\qbar^6\right).
\eq
The differential operator $L$ can be written in the $\qbar$-coordinate with $\theta_{\qbar}=\qbar \frac{d}{d\qbar}$ as
\bq
 L & = &
 \beta \theta_{\qbar} \frac{1}{\Yinvariant_{l-1}} \theta_{\qbar} \frac{1}{\Yinvariant_{l-2}} \theta_{\qbar} \frac{1}{\Yinvariant_{l-3}} \dots \frac{1}{\Yinvariant_3} \theta_{\qbar} \frac{1}{\Yinvariant_2} \theta_{\qbar} \frac{1}{\Yinvariant_1} \theta_{\qbar} 
 \frac{1}{\Frobeniusbasis_0},
\eq
where $\beta$ is a function of $\qbar$.
With $\Yinvariant_1=1$ and $\Yinvariant_j=\Yinvariant_{l-j}$ this simplifies to
\bq
\label{L_operator_q_coordinate}
 L & = &
 \beta \theta_{\qbar}^2 \frac{1}{\Yinvariant_{2}} \theta_{\qbar} \frac{1}{\Yinvariant_3} \dots \frac{1}{\Yinvariant_3} \theta_{\qbar} \frac{1}{\Yinvariant_2} \theta_{\qbar}^2 
 \frac{1}{\Frobeniusbasis_0}.
\eq
The operator
\bq
\label{special_local_normal_form}
 N\left(L\right)
 & = &
 \theta_{\qbar}^2 \frac{1}{\Yinvariant_{2}} \theta_{\qbar} \frac{1}{\Yinvariant_3} \dots \frac{1}{\Yinvariant_3} \theta_{\qbar} \frac{1}{\Yinvariant_2} \theta_{\qbar}^2 
\eq
is called the special local normal form of the operator $L$.

For the Picard--Fuchs operators $L^{(l,0)}$ we have with $\alpha$ given by eq.~(\ref{def_alpha_banana})
\bq
 \prod\limits_{j=1}^{l-1} \Yinvariant_j & = & \frac{J^{l-1}}{\Frobeniusbasis_0^2} \alpha
\eq
and
\bq
 \beta 
 & = &
 \frac{\alpha}{J \Frobeniusbasis_0}.
\eq
Let us discuss the form of eq.~(\ref{L_operator_q_coordinate}) 
and the special local normal form of eq.~(\ref{special_local_normal_form})
for the Picard--Fuchs operator $L^{(l,0)}$ of the banana integrals:
The left-multiplication with $\beta$ is of no particular importance, as we may always divide by this function.
The operator $L^{(l,0)}$ annihilates (by construction) 
the maximal cut of the $l$-loop banana integral $I_{1 \dots 1 1}$ in $D=2$ dimensions.
The special local normal form $N(L^{(l,0)})$ of this operator annihilates
$I_{1 \dots 1 1}/\Frobeniusbasis_0$ in $D=2$ dimensions, as the special local normal form
does not contain the right factor $1/\Frobeniusbasis_0$.
This suggests $f(\eps) \cdot I_{1 \dots 1 1}/\Frobeniusbasis_0$ as a master integral for the $\eps$-factorised basis,
where $f(\eps)$ is a function of $\eps$, but not of $y$.
In the next section we will see that the choice $f(\eps)=\eps^l$ leads to an $\eps$-factorised basis.
The special local normal forms of the operators $L^{(l,0)}$ up to six loops are
\bq
 N\left(L^{(1,0)}\right) & = & \theta_q,
 \nonumber \\
 N\left(L^{(2,0)}\right) & = & \theta_q^2,
 \nonumber \\
 N\left(L^{(3,0)}\right) & = & \theta_q^3,
 \nonumber \\
 N\left(L^{(4,0)}\right) & = & \theta_q^2 \frac{1}{\Yinvariant_2} \theta_q^2,
 \nonumber \\
 N\left(L^{(5,0)}\right) & = & \theta_q^2 \frac{1}{\Yinvariant_2} \theta_q \frac{1}{\Yinvariant_2} \theta_q^2,
 \nonumber \\
 N\left(L^{(6,0)}\right) & = & \theta_q^2 \frac{1}{\Yinvariant_2} \theta_q \frac{1}{\Yinvariant_3} \theta_q \frac{1}{\Yinvariant_2} \theta_q^2.
\eq
The non-trivial $\Yinvariant$-invariants enter only from $4$-loop onwards, i.e. 
for Calabi--Yau manifolds of dimension $3$ or higher.
The sequence of the special local normal forms is systematic, however knowing only the terms of loop order $l \le 3$ does not
allow us to deduce the general pattern.

\subsection{Calabi--Yau operators}

Apart from being self-dual and having a point with maximal unipotent monodromy with an integer local exponent,
the Picard--Fuchs operators $L^{(l,0)}$ of eq.~(\ref{def_Picard_Fuchs_eps_0}) 
have additional properties related to integral power series.
This brings us to the algebraic characterisation of Calabi--Yau operators.

A power series
\bq
 \sum\limits_{n=0}^\infty a_n y^n
\eq
is called $N$-integral, if there is a natural number $N$ such that $N^n a_n \in {\mathbb Z}$.
In other words, the substitution $y=N y'$ leads to a power series in the new variable $y'$ with integer coefficients.

A differential operator is called a Calabi--Yau operator if
\begin{enumerate}
\item $L$ is self-dual.
\item The point $y=0$ is a point of maximal unipotent monodromy and the local exponent at $y$ is an integer.
\item The holomorphic solution $\Frobeniusbasis_0$ as a power series in $y$ is $N$-integral.
\item The variable $\qbar$ as a power series in $y$ is $N$-integral.
\item All functions $(\alpha_1,\alpha_2,\dots,\alpha_{l-1})$ as power series in $y$ are $N$-integral.
\end{enumerate}
The Picard--Fuchs operators $L^{(l,0)}$ of eq.~(\ref{def_Picard_Fuchs_eps_0}) are Calabi--Yau operators.
We have seen examples for the conditions $(3)$ and $(5)$ 
in eq.~(\ref{examples_Frobenius_0}) and eq.~(\ref{examples_structure_series}), respectively.
Examples for condition $(4)$ are obtained from eq.~(\ref{examples_y_qbar_expansion}) by reversion of the power series.


\section{The method}
\label{sect:method}

In this section we consider the family of the $l$-loop equal mass banana integrals.
As we are considering a fixed loop order, we drop in this section the superscript $(l)$.

We present the method to cast the differential equation for the $l$-loop banana integrals
into an $\eps$-factorised form. 
This is based on an ansatz, which we give in sub-section~\ref{sect_ansatz}.
The ansatz involves a priori unknown functions, which are determined from algebraic equations (see eq.~(\ref{cond_algebraic}))
and differential equations (see eq.~(\ref{cond_poles})).
In sub-section~\ref{sect_diff_eq} we present the final differential equation in $\eps$-factorised form 
and introduce iterated integrals.
In addition to the differential equation we need boundary values, which we give in sub-section~\ref{sect_diff_boundary}.

\subsection{The ansatz for the master integrals}
\label{sect_ansatz}

In this sub-section we construct the master integrals
\bq
 M & = & \left(M_{0}, M_{1}, \dots, M_{l}\right)^T,
\eq
which put the differential equation into an $\eps$-factorised form.

The master integral $M_{0}$ is related to the tadpole integral and is rather simple.
We set
\bq
 M_{0}
 & = & 
 \eps^l I_{1 \dots 1 0}
 \; = \; 
  \left[ e^{\gamma_E \eps} \Gamma\left(1+\eps\right) \right]^l.
\eq
For the master integral $M_{1}$ we set
\bq
 M_{1}
 & = & 
 \frac{\eps^l}{\Frobeniusbasis_0} I_{1 \dots 1 1}.
\eq
For the master integrals $M_2-M_l$ we make the ansatz
\bq
 M_{j}
 & = &
 \frac{1}{\Yinvariant_{j-1}} 
 \left[  
  \frac{J}{\eps} \frac{d}{dy} M_{j-1}
  - \sum\limits_{k=1}^{j-1} F_{(j-1)k} M_{k}
 \right],
\eq
with a priori unknown functions $F_{ij}$, which depend on $y$ (or $\tau$), but not on $\eps$.
The function $J$ denotes the Jacobian, defined in eq.~(\ref{def_Jacobian}).
The functions $\Yinvariant_j$ have been defined in section~\ref{sect:calabi_yau}.
Note that we have 
\bq
 \Yinvariant_1 \; = \; 1
 & \mbox{and} &
 \Yinvariant_{l-i} \; = \; \Yinvariant_i.
\eq
From this ansatz it follows immediately that the first $l$ rows of the differential equation are
\bq
 J \frac{d}{dy} M
 & = &
 \eps
 \left( \begin{array}{cccccccc}
 0 & 0 & 0 & 0 & 0 & \dots & 0 & 0 \\
 0 & F_{11} & 1 & 0 & 0 & & 0 & 0 \\
 0 & F_{21} & F_{22} & \Yinvariant_2 & 0 & & 0 & 0 \\
 0 & F_{31} & F_{32} & F_{33} & \Yinvariant_3 & & 0 & 0 \\
 \vdots & & & & & \ddots & & \vdots \\
 0 & F_{(l-2) 1} & F_{(l-2) 2} & F_{(l-2) 3} & F_{(l-2) 4} & \dots & \Yinvariant_{l-2} & 0 \\
 0 & F_{(l-1) 1} & F_{(l-1) 2} & F_{(l-1) 3} & F_{(l-1) 4} & \dots & F_{(l-1) (l-1)} & 1 \\
 \ast & \ast & \ast & \ast & \ast & \dots & \ast & \ast \\
 \end{array} \right) M.
\eq
The first $l$ rows are in an $\eps$-factorised form.
It remains to choose the functions $F_{ij}$ such that the $(l+1)$-th row is in $\eps$-factorised form as well.
Let us write
\bq
 J \frac{d}{dy} M
 & = &
 A M,
\eq
where we label the entries $A_{ij}$ of the matrix $A$ with indices from the range $\{0,1,\dots,l\}$.
We have
\bq
 A_{ij} & = & F_{ij}
 \;\;\; \mbox{for} \;\; i \in \{1,\dots,l-1\} \;\; \mbox{and} \;\; 1 \le j \le i.
\eq
It turns out that $A_{l 0}$ is always $\eps$-factorised and given by
\bq
 A_{l 0}
 & = &
 \eps \left(-1\right)^l \left(l+1\right)! \frac{\Frobeniusbasis_0 J}{y^2}.
\eq
The entries $A_{l k}$ for $k \in \{1,\dots,l\}$ are of the form
\bq
 A_{l k}
 & = &
 \sum\limits_{j=k-l}^1 A_{l k}^{(j)} \eps^j,
\eq
where the $A_{l k}^{(j)}$ are independent of $\eps$.
We require that the $A_{l k}^{(j)}$ with $j<1$ vanish:
\bq
\label{cond_poles}
 A_{l k}^{(j)} & = & 0
 \;\;\; \mbox{for} \;\; j < 1.
\eq
This leads to differential equations for the unknown functions $F_{ij}$.
Actually, we may impose a stronger constraint:
Self-duality allows us to impose the conditions
\bq
\label{cond_self_duality}
 A_{i j} & = & A_{(l+1-j) (l+1-i)}
 \;\;\; \mbox{for} \;\;
 i,j \in \{1,\dots,l\}.
\eq
First of all, this equation eliminates directly some of the $F_{ij}$, as we have
\bq
\label{cond_trivial}
 F_{i j} & = & F_{(l+1-j) (l+1-i)}
 \;\;\; \mbox{for} \;\;
 i \in \{2,\dots,l-1\}  \;\; \mbox{and} \;\; j \le i.
\eq
We call eq.~(\ref{cond_trivial}) the trivial equations.
The trivial equations reduce the number of unknown functions.
Secondly, self-duality implies the differential equations of eq.~(\ref{cond_poles}).
Thirdly, we get from the $\eps^1$-term of the last row the algebraic equations
\bq
\label{cond_algebraic}
 A_{l k}^{(1)} - F_{(l+1-k) 1} & = & 0.
\eq
It is advantageous to use first the trivial equations of eq.~(\ref{cond_trivial}), then to solve all algebraic equations of eq.~(\ref{cond_algebraic})
and finally the differential equations of eq.~(\ref{cond_poles}).

\subsection{The differential equation}
\label{sect_diff_eq}

Having determined the $F_{ij}$ it is convenient to change the notation and write
\bq
\label{A_f_notation}
 A
 & = &
 \eps
 \left( \begin{array}{cccccccc}
 0 & 0 & 0 & 0 & 0 & \dots & 0 & 0 \\
 0 & f_{2,1} & f_{0,2} & 0 & 0 & & 0 & 0 \\
 0 & f_{4,1} & f_{2,2} & f_{0,3} & 0 & & 0 & 0 \\
 0 & f_{6,1} & f_{4,2} & f_{2,3} & f_{0,4} & & 0 & 0 \\
 \vdots & & & & & \ddots & & \vdots \\
 0 & f_{2(l-2),1} & f_{2(l-3),2} & f_{2(l-4),3} & f_{2(l-5),4} & \dots & f_{0,(l-1)} & 0 \\
 0 & f_{2(l-1),1} & f_{2(l-2),2} & f_{2(l-3),3} & f_{2(l-4),4} & \dots & f_{2,(l-1)} & f_{0,l} \\
 f_{l+1,0} & f_{2l,1} & f_{2(l-1),2} & f_{2(l-2),3} & f_{2(l-3),4} & \dots & f_{4,(l-1)} & f_{2,l} \\
 \end{array} \right).
\eq
The symmetry of eq.~(\ref{cond_self_duality}) translates to
\bq
 f_{2i,j} & = & f_{2i,l+2-i-j},
 \;\;\; \mbox{for} \;\;
 i,j \in \{1,\dots,l\}.
\eq
We have
\bq
 f_{0,j} & = & \Yinvariant_{j-1},
 \nonumber \\
 f_{2(i+1-j),j} & = & F_{i j},
 \;\;\; \mbox{for} \;\;
 i \in \{1,\dots,l-1\}.
\eq
The motivation for this change of notation is the following:
In the two-loop case the $f_{i,j}$ are modular forms.
The first index $i$ corresponds to the modular weight, the second index $j$ distinguishes different
modular forms of the same modular weight.
This generalises to the $l$-loop case: We associate the (automorphic) weight $i$ to $f_{i,j}$.
The second index $j$ distinguishes different functions of the same weight $i$.
The weight counting assigns weight $(l-1)$ to $\Frobeniusbasis_0$ and weight $2$ to $J$.
The functions $\Yinvariant_j$ have weight zero.

We set
\bq
 \omega_{i,j} & = & 2 \pi i \; f_{i,j}\left(\tau\right) \; d\tau.
\eq
The differential equation reads then
\bq
 d M & = & \eps \Omega M,
\eq
where
\bq
\label{def_Omega}
 \Omega
 & = &
 \left( \begin{array}{cccccccc}
 0 & 0 & 0 & 0 & 0 & \dots & 0 & 0 \\
 0 & \omega_{2,1} & \omega_{0,2} & 0 & 0 & & 0 & 0 \\
 0 & \omega_{4,1} & \omega_{2,2} & \omega_{0,3} & 0 & & 0 & 0 \\
 0 & \omega_{6,1} & \omega_{4,2} & \omega_{2,3} & \omega_{0,4} & & 0 & 0 \\
 \vdots & & & & & \ddots & & \vdots \\
 0 & \omega_{2(l-2),1} & \omega_{2(l-3),2} & \omega_{2(l-4),3} & \omega_{2(l-5),4} & \dots & \omega_{0,(l-1)} & 0 \\
 0 & \omega_{2(l-1),1} & \omega_{2(l-2),2} & \omega_{2(l-3),3} & \omega_{2(l-4),4} & \dots & \omega_{2,(l-1)} & \omega_{0,l} \\
 \omega_{l+1,0} & \omega_{2l,1} & \omega_{2(l-1),2} & \omega_{2(l-2),3} & \omega_{2(l-3),4} & \dots & \omega_{4,(l-1)} & \omega_{2,l} \\
 \end{array} \right).
\eq
This differential equation can be solved systematically order-by-order in $\eps$ in terms of iterated integrals \cite{Chen}.
We define the $n$-fold iterated integral from $\tau_0$ to $\tau$ by
\begin{align}
I\left(\omega_{i_1,j_1},\omega_{i_2,j_2},...,\omega_{i_n,j_n};\tau,\tau_0\right)
 & =
 \left(2 \pi i \right)^n
 \int\limits_{\tau_0}^{\tau} d\tau_1
 \int\limits_{\tau_0}^{\tau_1} d\tau_2
 ...
 \int\limits_{\tau_0}^{\tau_{n-1}} d\tau_n
 \;
 f_{i_1,j_1}\left(\tau_1\right)
 f_{i_2,j_2}\left(\tau_2\right)
 ...
 f_{i_n,j_n}\left(\tau_n\right).
\end{align}
With $\qbar=\exp(2\pi i \tau)$ we may equally well write
\bq
I\left(\omega_{i_1,j_1},\omega_{i_2,j_2},...,\omega_{i_n,j_n};\tau,\tau_0\right)
 & = &
 \int\limits_{\qbar_0}^{\qbar} \frac{d\qbar_1}{\qbar_1}
 \int\limits_{\qbar_0}^{\qbar_1} \frac{d\qbar_2}{\qbar_2}
 ...
 \int\limits_{\qbar_0}^{\qbar_{n-1}} \frac{d\qbar_n}{\qbar_n}
 \;
 f_{i_1,j_1}\left(\tau_1\right)
 f_{i_2,j_2}\left(\tau_2\right)
 ...
 f_{i_n,j_n}\left(\tau_n\right),
 \nonumber \\
 & & \tau_j = \frac{1}{2\pi i} \ln \qbar_j.
\eq
Our standard choice for the base point $\tau_0$ will be $\tau_0 = i \infty$, corresponding to $q_0=0$.
If $f_{i_n,j_n}(\tau)$ does not vanish at $\tau=i\infty$ 
we employ the standard ``trailing zero'' or ``tangential base point'' regularisation \cite{Brown:2014aa,Adams:2017ejb,Walden:2020odh}:
We first take $\qbar_0$ to have a small non-zero value.
The integration will produce terms with $\ln(\qbar_0)$.
Let $R$ be the operator, which removes all $\ln(\qbar_0)$-terms.
After these terms have been removed, we may take the limit $\qbar_0\rightarrow 0$.
With this regularisation we set
\begin{align}
I\left(\omega_{i_1,j_1},\omega_{i_2,j_2},...,\omega_{i_n,j_n};\tau\right)
 & =
 \lim\limits_{\qbar_0\rightarrow 0}
 R \left[
 \int\limits_{\qbar_0}^{\qbar} \frac{d\qbar_1}{\qbar_1}
 \int\limits_{\qbar_0}^{\qbar_1} \frac{d\qbar_2}{\qbar_2}
 ...
 \int\limits_{\qbar_0}^{\qbar_{n-1}} \frac{d\qbar_n}{\qbar_n}
 \;
 f_{i_1,j_1}\left(\tau_1\right)
 f_{i_2,j_2}\left(\tau_2\right)
 ...
 f_{i_n,j_n}\left(\tau_n\right)
 \right].
\end{align}
As the last argument of all iterated integrals will always be $\tau$ and as it is sufficient to denote the $f_{i,j}$'
instead of the $\omega_{i,j}$ we introduce the short-hand notation
\bq
 I\left(f_{i_1,j_1},f_{i_2,j_2},...,f_{i_n,j_n}\right)
 & = &
 I\left(\omega_{i_1,j_1},\omega_{i_2,j_2},...,\omega_{i_n,j_n};\tau\right).
\eq
The entries on the diagonal of the matrix $\Omega$ are
$\omega_{2,1}, \dots, \omega_{2,l}$.
We may separate them into a common dlog-form $\omega_2^{\mathrm{mpl}}$ and a remainder $\tilde{\omega}_{2,j}$ as
\bq
\label{split_mpl_remainder}
 \omega_{2,j} & = & \omega_2^{\mathrm{mpl}} + \tilde{\omega}_{2,j},
\eq
where
\bq
 f_2^{\mathrm{mpl}} & = & J \left[ \frac{\left(l+1\right)}{2} \frac{1}{y} - \sum\limits_{a \in S^{(l)}} \frac{a}{\left(1+ay\right)} \right],
 \nonumber \\
 \omega_2^{\mathrm{mpl}}
 & = &
 2 \pi i \; f_2^{\mathrm{mpl}} \; d\tau
 \; = \; 
 \frac{\left(l+1\right)}{2} d\ln\left(y\right) - \sum\limits_{a \in S^{(l)}} d\ln\left(1+ay\right).
\eq
This notation is convenient, as one of the algebraic equations turns into
\bq
 \sum\limits_{j=1}^l \tilde{\omega}_{2,j}
 & = & 0.
\eq
This equation together with eq.~(\ref{cond_self_duality}) implies that the 
$\tilde{\omega}_{2,j}$ can only be non-zero for $l \ge 3$.
For $l=2$ we have with eq.~(\ref{cond_self_duality})
\bq
 \tilde{\omega}_{2,1} + \tilde{\omega}_{2,2} \; = \; 2 \tilde{\omega}_{2,1} \; = \; 0.
\eq

\subsection{The boundary values}
\label{sect_diff_boundary}

With the differential equation in $\eps$-factorised form at hand we only need the boundary values as additional input.
We choose $y=0$ as boundary point.
It is sufficient to know the boundary value of $M^{(l)}_1$, the boundary values of the other master integrals
$M^{(l)}_k$ with $k>1$ follow from the higher orders in the dimensional regularisation parameter $\eps$ of $M^{(l)}_1$.
For $M^{(l)}_1$ we need the constant term and all logarithms $\ln(y)$.
The boundary value is easily obtained with the help of the Mellin--Barnes technique.
The calculation follows the lines of \cite{Broedel:2019kmn,Bonisch:2021yfw,Pogel:2022yat}.
The result is 
\bq
\label{boundary_value}
 \left. M^{(l)}_1 \right|_{y \to 0}
 & = &
 e^{l \eps \gamma_E}
 \left(l+1\right)
 \sum\limits_{j=0}^{l}
  \left( \begin{array}{c} l \\ j \\ \end{array}\right)
  \left(-1\right)^j
  y^{j\eps}
  \frac{\Gamma\left(1+\eps\right)^{l-j}\Gamma\left(1-\eps\right)^{1+j}\Gamma\left(1+j\eps\right)}{\Gamma\left(1-\left(j+1\right)\eps\right)}.
\eq


\section{The degenerate cases of one loop and zero loops}
\label{sect:degenerated_cases}

It is worth discussing the one-loop case and the zero-loop case from the view point of the general $l$-loop case.
In particular we are interested in the change of variables from $y$ to $\tau$ (or $\qbar$).
It turns out that these can be extrapolated to $l=1$ and $l=0$.

\subsection{The one-loop case}

The second graph polynomial ${\mathcal F}$ is given in the one-loop case by
\bq
 {\mathcal F} = - a_1 a_2 x + \left( a_1 + a_2 \right)^2.
\eq
The zero set $X$ of ${\mathcal F}=0$ in ${\mathbb C} {\mathbb P}^{1}$ consists for generic $x$ of two points
\bq
 \left[ \frac{1}{2}\left(x-2-\sqrt{-x\left(4-x\right)}\right) : 1 \right],
 & &
 \left[ \frac{1}{2}\left(x-2+\sqrt{-x\left(4-x\right)}\right) : 1 \right].
\eq
It is a disconnected zero-dimensional manifold with two connected components.
We therefore obtain the Hodge number $h^{0,0}=2$.
It is not a zero-dimensional Calabi--Yau manifold, as for a Calabi--Yau manifold we would have $h^{0,0}=1$.

The Picard--Fuchs operator $L^{(1,0)}$ is given by
\bq
 L^{(1,0)} 
 & = &
 \frac{d}{dy} - \frac{1}{y} + \frac{1}{2} \frac{4}{\left(1+4y\right)}.
\eq
This is a first-order differential operator and there is one independent solution $\Frobeniusbasis_0$, 
given by eq.~(\ref{Frobenius_series_holomorphic}).
Eq.~(\ref{def_a_l_0_n}) reduces to 
\bq
 a_{0,n}
 & = &
 \left(-1\right)^n
 \left( \begin{array}{c} 2 n \\ n \\ \end{array} \right)
\eq
and we obtain
\bq
 \Frobeniusbasis_0
 \; = \;
 \frac{y}{\sqrt{1+4y}}.
\eq
The general formulae of eq.~(\ref{Frobenius_series_single_log}) and eq.~(\ref{def_a_l_1_n}) 
also make sense for $l=1$, yielding
\bq
 a_{1,n}
 & = &
 2 \left(-1\right)^n
 \left( \begin{array}{c} 2 n \\ n \\ \end{array} \right)
 \left[ S_1\left(2n\right) - S_1\left(n\right) \right]
\eq
and 
\bq
 \Frobeniusbasis_1
 & = &
 \frac{1}{2\pi i} \ln\left(\frac{\sqrt{1+4y}-1}{\sqrt{1+4y}+1} \right)  \Frobeniusbasis_0.
\eq
We emphasize that $\Frobeniusbasis_1$ is not a solution of $L^{(1,0)}\Frobeniusbasis=0$,
it is the extrapolation of eq.~(\ref{Frobenius_series_single_log}) and eq.~(\ref{def_a_l_1_n}) to $l=1$.

We may therefore define also for $l=1$ a change of variables from $y$ to $\tau$ (or $\qbar$) 
as we did for $l \ge 2$.
We obtain
\bq
 \tau \; = \; \frac{1}{2\pi i} \ln\left(\frac{\sqrt{1+4y}-1}{\sqrt{1+4y}+1} \right),
 & &
 \qbar \; = \; \frac{\sqrt{1+4y}-1}{\sqrt{1+4y}+1},
 \;\;\;\;\;\;
 y \; = \; \frac{\qbar}{\left(1-\qbar\right)^2}.
\eq
We note that this change of variables from $y$ to $\qbar$ rationalises the square root $\sqrt{1+4y}$:
\bq
 \sqrt{1+4y} & = & \frac{1+\qbar}{1-\qbar}.
\eq

\subsection{The zero-loop case}

From eq.~(\ref{def_graph_polynomials}) be obtain the second graph polynomial for zero loops as
\bq
 {\mathcal F} & = &
 a_1 \left(1-x\right).
\eq
For generic $x$ the zero set $X$ of ${\mathcal F}=0$ in ${\mathbb C} {\mathbb P}^{0}$ 
is the empty set
\bq
 X & = & \emptyset.
\eq
The Picard--Fuchs operator would be a differential operator of order zero and normalising the leading coefficient to one yields $L^{(0,0)}=1$.
The equation $L^{(0,0)} \Frobeniusbasis = 0$
has a zero-dimensional solution space, consisting of the trivial solution $\Frobeniusbasis=0$ only.
However, eq.~(\ref{Frobenius_series_holomorphic}), eq.~(\ref{def_a_l_0_n}), eq.~(\ref{Frobenius_series_single_log}) and eq.~(\ref{def_a_l_1_n}) also make sense for $l=0$, yielding
\bq
 a_{0,n}
 \; = \;
 \left(-1\right)^n
 & \mbox{and} &
 a_{1,n}
 \; = \;
 0
\eq
as well as
\bq
 \Frobeniusbasis_0
 \; = \;
 \frac{y}{1+y}
 & \mbox{and} &
 \Frobeniusbasis_1
 \; = \;
 \frac{\ln y}{\left(2\pi i\right)} \Frobeniusbasis_0.
\eq
We emphasize that $\Frobeniusbasis_0$ and $\Frobeniusbasis_1$ are not solutions of $L^{(0,0)}\Frobeniusbasis=0$,
they are the extrapolation of eq.~(\ref{Frobenius_series_holomorphic}), eq.~(\ref{def_a_l_0_n}), eq.~(\ref{Frobenius_series_single_log}) and eq.~(\ref{def_a_l_1_n})
to $l=0$.
From $\Frobeniusbasis_0$ and $\Frobeniusbasis_1$ we obtain the change of variables
\bq
 \tau \; = \; \frac{\ln y}{\left(2\pi i\right)},
 & &
 \qbar \; = \; y.
\eq
We find that at zero loops the change of variables from $y$ to $\qbar$ is the identity map.
 

\section{Example: 5 loops}
\label{sect:5_loops}

In this section we discuss the equal-mass five-loop banana integral.
The $\eps$-factorised differential equations for the equal-mass banana integrals with up to four loops
have already been discussed in the literature:
The four-loop case has been discussed in ref.~\cite{Pogel:2022ken}, the three-loop case in ref.~\cite{Pogel:2022yat},
the two-loop case in ref.~\cite{Adams:2018yfj}.
The one-loop case is rather trivial, a pedagogical discussion can be found in \cite{Weinzierl:2022eaz}.
The five-loop case is therefore the first case, where our method yields new results beyond the current state-of-the-art.

\begin{figure}
\begin{center}
\includegraphics[scale=0.5]{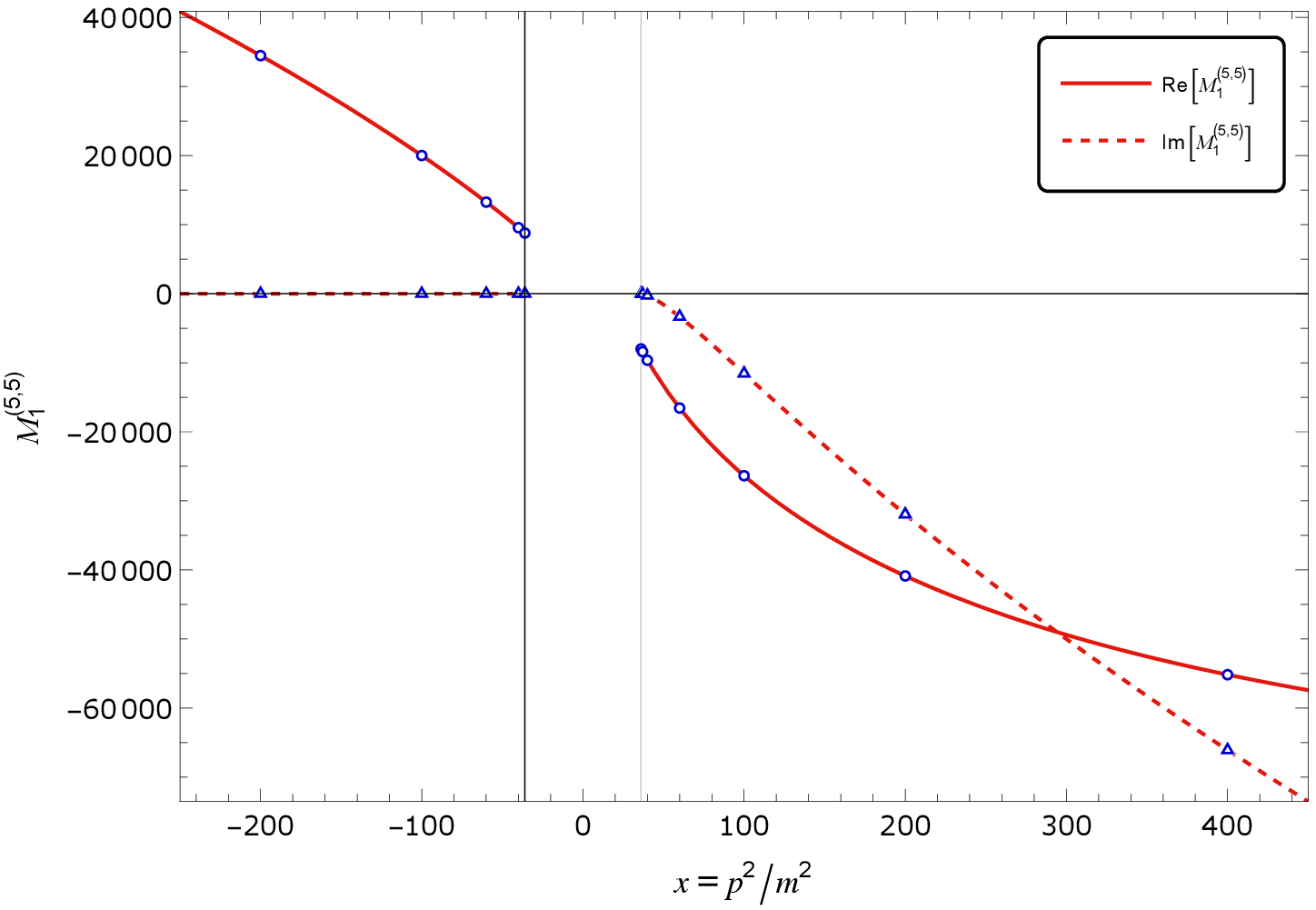}
\hspace*{5mm}
\includegraphics[scale=0.5]{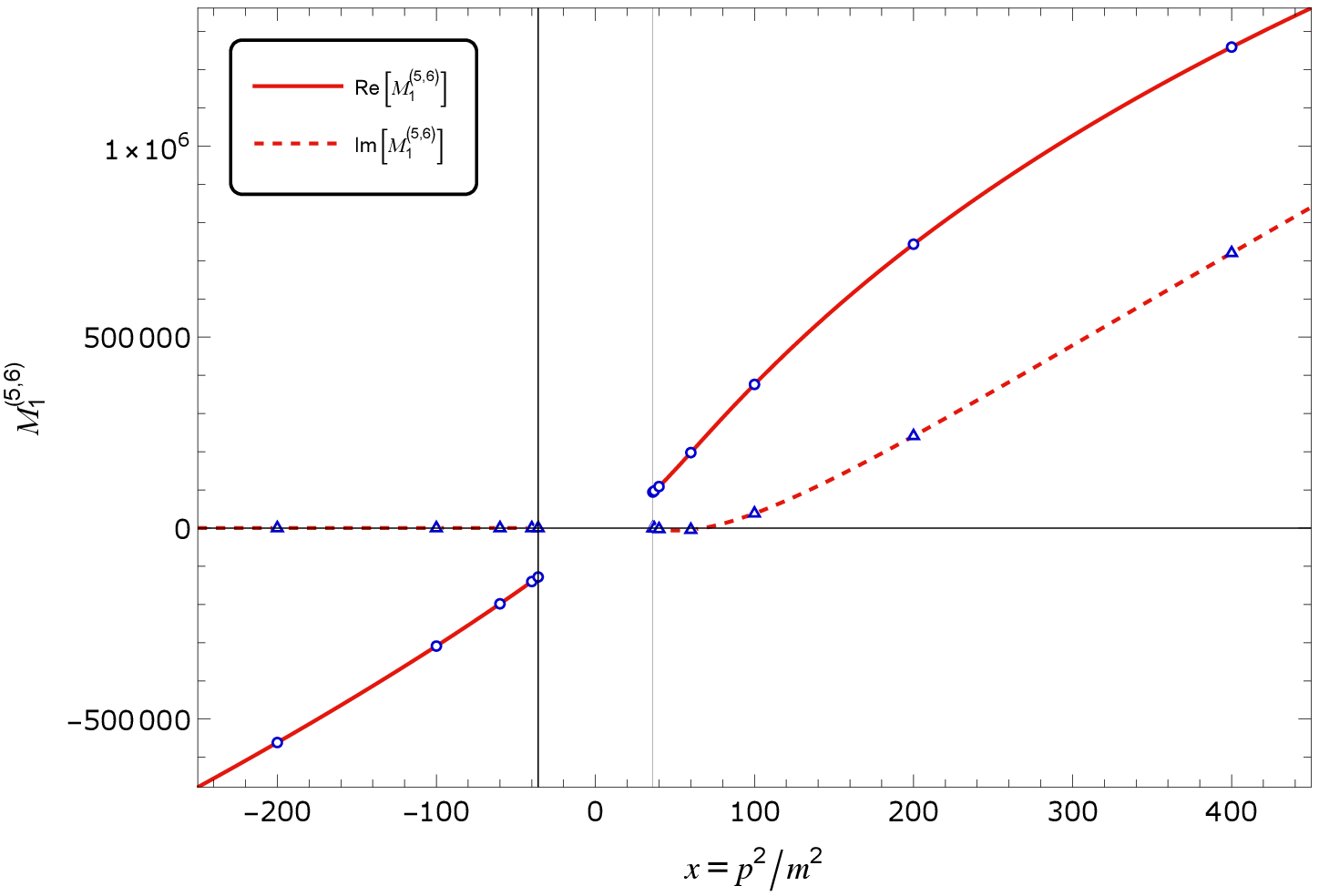}
\end{center}
\caption{
Comparison of our result for $M_1^{(5,5)}$ and $M_1^{(5,6)}$ at five loops with numerical results from \tt{pySecDec}.
}
\label{fig_result_5_loops}
\end{figure}

Our ansatz at five loops reads
\bq
 M_{0} 
 & = &
 \eps^5 I_{111110},
 \nonumber \\
 M_{1}
 & = & 
 \frac{\eps^5}{\Frobeniusbasis_0} I_{111111},
 \nonumber \\
 M_{2}
 & = & 
  \frac{J}{\eps} \frac{d}{dy} M_{1}
  - F_{11} M_{1},
 \nonumber \\
 M_{3}
 & = & 
 \frac{1}{\Yinvariant_{2}} 
 \left[  
  \frac{J}{\eps} \frac{d}{dy} M_{2}
  - F_{21} M_{1}
  - F_{22} M_{2}
 \right],
 \nonumber \\
 M_{4}
 & = & 
 \frac{1}{\Yinvariant_{2}} 
 \left[  
  \frac{J}{\eps} \frac{d}{dy} M_{3}
  - F_{31} M_{1}
  - F_{32} M_{2}
  - F_{33} M_{3}
 \right],
 \nonumber \\
 M_{5}
 & = & 
  \frac{J}{\eps} \frac{d}{dy} M_{4}
  - F_{41} M_{1}
  - F_{42} M_{2}
  - F_{32} M_{3}
  - F_{22} M_{4}.
\eq
Here we used already $\Yinvariant_1=\Yinvariant_4=1$, $\Yinvariant_3=\Yinvariant_2$ and the trivial equations $F_{44}=F_{22}$ and $F_{43}=F_{32}$.
There are four algebraic equations, which can be used to eliminate
$F_{33}$, $F_{32}$, $F_{42}$ and $F_{41}$.
If we write
\bq
 F_{11} \; = \;  f_2^{\mathrm{mpl}} + \tilde{f}_{2,1},
 \; \;\;
 F_{22} \; = \;  f_2^{\mathrm{mpl}} + \tilde{f}_{2,2},
 \; \;\;
 F_{33} \; = \;  f_2^{\mathrm{mpl}} + \tilde{f}_{2,3},
\eq
as we did in eq.~(\ref{split_mpl_remainder}), one of the algebraic equations equals
\bq
 2 \tilde{f}_{2,1} + 2 \tilde{f}_{2,2} + \tilde{f}_{2,3} & = & 0.
\eq
This leaves the functions $F_{11}$, $F_{21}$, $F_{22}$ and $F_{31}$, which are determined by the differential equations
of eq.~(\ref{cond_poles}).
The differential equations of eq.~(\ref{cond_poles}) lead also to a fourth order non-linear differential equation
for $\Frobeniusbasis_0$.
We recall that $\Frobeniusbasis_0$ is by definition the solution of a fifth order linear differential equation, holomorphic
at $y=0$.
It is easily checked that $\Frobeniusbasis_0$ fulfils the fourth order non-linear differential equation.
We will discuss this in more detail in section~\ref{sect:relations}.
Solving all equations we obtain with the notation as in eq.~(\ref{A_f_notation}) for the first few terms of the $\qbar$-expansion
\bq
 f_{0,2} & = & 1,
 \\
 f_{0,3} 
 & = & 
 1 - 2 \qbar + 46 \qbar^2 - 1010 \qbar^3 + 21550 \qbar^4 - 463502 \qbar^5+\mathcal{O}(\qbar^6),
 \nonumber \\
 f_{2,1}
 & = &
 \frac{5}{2} - 10 \qbar + 50 \qbar^2 - 1090 \qbar^3 + 18770 \qbar^4 - 360310 \qbar^5+\mathcal{O}(\qbar^6),
 \nonumber \\
 f_{2,2}
 & = &
 \frac{5}{2} - 32 \qbar + 616 \qbar^2 - 14720 \qbar^3 + 338440 \qbar^4 - 7750832 \qbar^5+\mathcal{O}(\qbar^6),
 \nonumber \\
 f_{2,3}
 & = &
 5 - 46 \qbar + 1058 \qbar^2 - 27910 \qbar^3 + 703970 \qbar^4 - 17298946 \qbar^5+\mathcal{O}(\qbar^6),
 \nonumber \\
 f_{4,1}
 & = &
 -\frac{1}{2} \left( 105 \qbar - 3075 \qbar^2 + 79305 \qbar^3 - 2011395 \qbar^4 + 49317855 \qbar^5 \right) + \mathcal{O}(\qbar^6),
 \nonumber \\
 f_{4,2}
 & = &
 \frac{5}{4} - 94 \qbar + 3842 \qbar^2 - 133870 \qbar^3 + 4204610 \qbar^4 - 120866194 \qbar^5+\mathcal{O}(\qbar^6),
 \nonumber \\
 f_{6,0}
 & = &
 -720\left(1 + 4 \qbar - 44 \qbar^2 + 364 \qbar^3 - 5804 \qbar^4 + 95404 \qbar^5\right) +\mathcal{O}(\qbar^6), 
 \nonumber \\
 f_{6,1}
 & = &
 -\frac{1}{4} \left( 105 \qbar - 14715 \qbar^2 + 787425 \qbar^3 - 30754395 \qbar^4 + 1020051855 \qbar^5 \right) +\mathcal{O}(\qbar^6),
 \nonumber \\
 f_{6,2}
 & = &
 -\frac{25}{4} + 128 \qbar - 304 \qbar^2 - 168640 \qbar^3 + 10537040 \qbar^4 - 438453472 \qbar^5+\mathcal{O}(\qbar^6),
 \nonumber \\
 f_{8,1}
 & = &
 -\frac{1}{32} \left(9 - 2520 \qbar + 169080 \qbar^2 - 1366200 \qbar^3 - 261503880 \qbar^4 +  18190697880 \qbar^5 \right) 
 \nonumber \\
 & &
 +\mathcal{O}(\qbar^6),
 \nonumber \\
 f_{10,1}
 & = &
  \frac{1}{32}\left(45 - 4860 \qbar - 42660 \qbar^2 + 7549380 \qbar^3 - 81509220 \qbar^4 - 13609216260 \qbar^5\right) +\mathcal{O}(\qbar^6).
 \nonumber 
\eq
We then solve the differential equation for the master integrals $M_0, M_1, \dots, M_5$
with the boundary condition given by eq.~(\ref{boundary_value}).
The $\eps$-expansion of the master integral $M_1$ starts at order $\eps^5$:
\bq
 M_1 & = & 
 \eps^5 M_1^{(5,5)} + \eps^6 M_1^{(5,6)}
 + {\mathcal O}\left(\eps^7\right).
\eq
The first term in the $\eps$-expansion is given by
\bq
 M_1^{(5,5)}
 & = &
 288\zeta_5 + 480 \zeta_3 I\left(1,\Yinvariant_2\right) + I\left(1,\Yinvariant_2,\Yinvariant_2,1,f_{6,0}\right).
\eq
The first few terms of the $\qbar$-expansion of $M_1^{(5,5)}$ 
read with $L_{\qbar}=\ln(\qbar)$
\bq
 M_1^{(5,5)}
 & = & 
 288 \zeta_5 + 240 \zeta_3 L_{\qbar}^2 - 6 L_{\qbar}^5 
 +240 {\qbar} \left(-4 \zeta_3 + L_{\qbar}^3-3 L_{\qbar}^2 \right)
 \nonumber \\
 & &
 -30 {\qbar}^2 \left(-184 \zeta_3+46 L_{\qbar}^3-57 L_{\qbar}^2-48 L_{\qbar}+30\right)
 + {\mathcal O}\left(\qbar^3\right).
\eq
In fig.~\ref{fig_result_5_loops} we plot the results for $M_1^{(5,5)}$ and $M_1^{(5,6)}$ for $|x|>36$.
We also plotted the results from the program $\verb|pySecDec|$ \cite{Borowka:2017idc}.
We observe excellent agreement.


\section{Example: 6 loops}
\label{sect:6_loops}

\begin{figure}
\begin{center}
\includegraphics[scale=0.5]{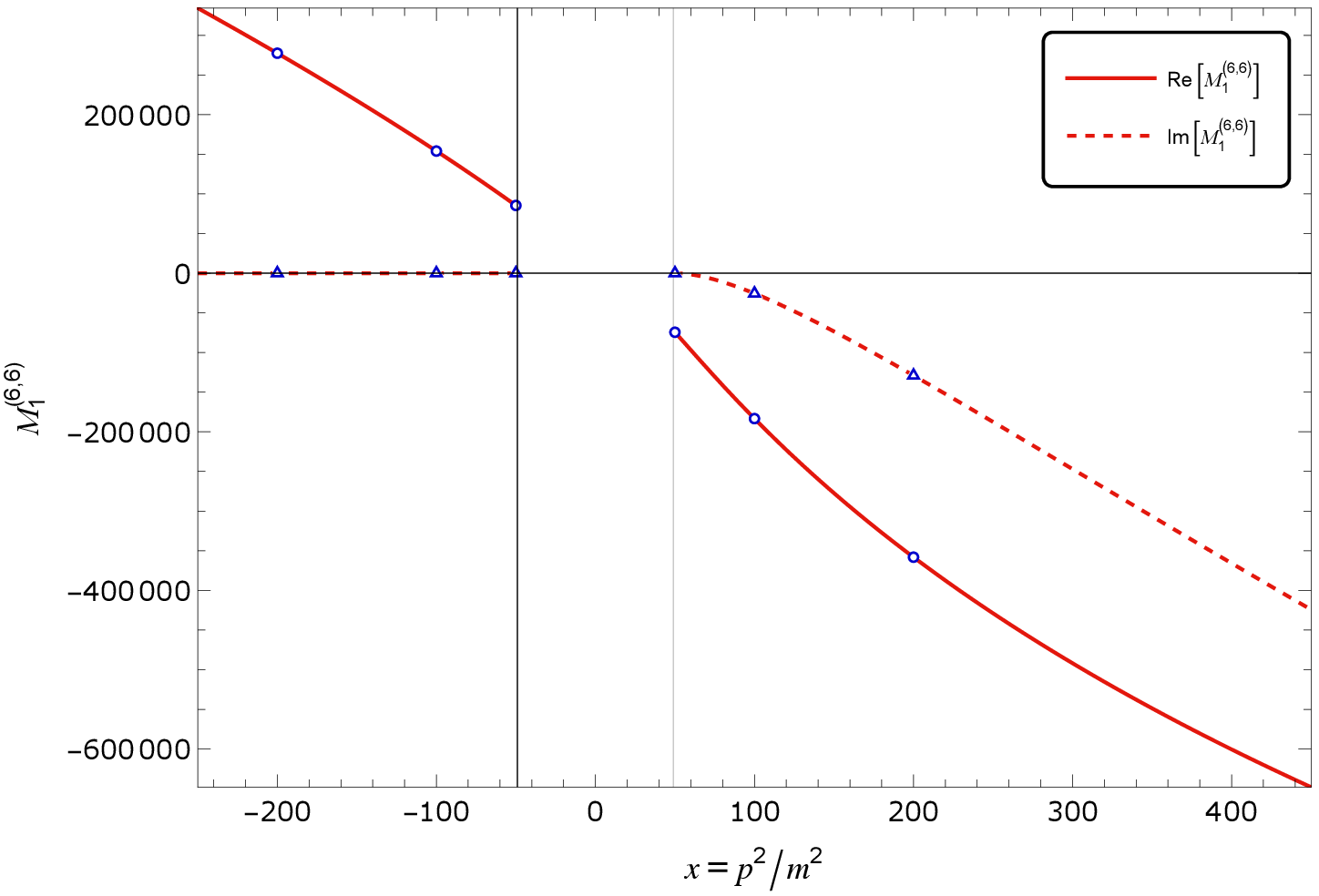}
\hspace*{5mm}
\includegraphics[scale=0.5]{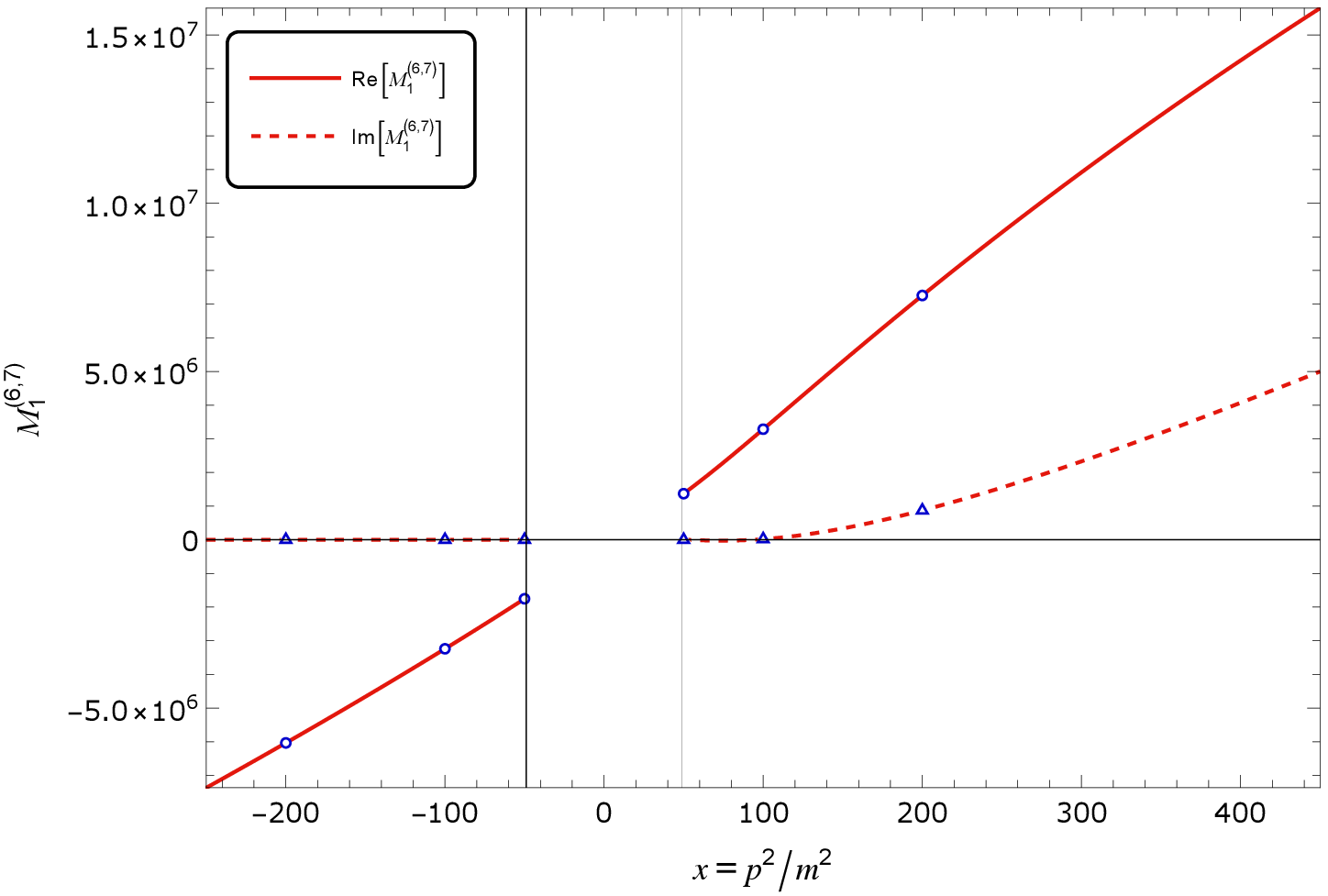}
\end{center}
\caption{
Comparison of our result for $M_1^{(6,6)}$ and $M_1^{(6,7)}$ at six loops with numerical results from \tt{pySecDec}.
}
\label{fig_result_6_loops}
\end{figure}

The six-loop case is of interest, because it is the first example where two non-trivial $\Yinvariant$-invariants appear.
At six loops the $\Yinvariant$-invariants $\Yinvariant_2$ and $\Yinvariant_3$ enter.
Our ansatz reads
\bq
 M_{0} 
 & = &
 \eps^6 I_{1111110},
 \nonumber \\
 M_{1}
 & = & 
 \frac{\eps^6}{\Frobeniusbasis_0} I_{1111111},
 \nonumber \\
 M_{2}
 & = & 
  \frac{J}{\eps} \frac{d}{dy} M_{1}
  - F_{11} M_{1},
 \nonumber \\
 M_{3}
 & = & 
 \frac{1}{\Yinvariant_{2}} 
 \left[  
  \frac{J}{\eps} \frac{d}{dy} M_{2}
  - F_{21} M_{1}
  - F_{22} M_{2}
 \right],
 \nonumber \\
 M_{4}
 & = & 
 \frac{1}{\Yinvariant_{3}} 
 \left[  
  \frac{J}{\eps} \frac{d}{dy} M_{3}
  - F_{31} M_{1}
  - F_{32} M_{2}
  - F_{33} M_{3}
 \right],
 \nonumber \\
 M_{5}
 & = & 
 \frac{1}{\Yinvariant_{2}} 
 \left[  
  \frac{J}{\eps} \frac{d}{dy} M_{4}
  - F_{41} M_{1}
  - F_{42} M_{2}
  - F_{43} M_{3}
  - F_{33} M_{4}
 \right],
 \nonumber \\
 M_{6}
 & = & 
  \frac{J}{\eps} \frac{d}{dy} M_{5}
  - F_{51} M_{1}
  - F_{52} M_{2}
  - F_{42} M_{3}
  - F_{32} M_{4}
  - F_{22} M_{5}.
\eq
Here we used already $\Yinvariant_1=\Yinvariant_5=1$, $\Yinvariant_4=\Yinvariant_2$ and the trivial equations 
\bq
 F_{44} \;=\; F_{33},
 \;\;\;
 F_{53} \;=\; F_{42},
 \;\;\;
 F_{54} \;=\; F_{32},
 \;\;\;
 F_{55} \;=\; F_{22}.
\eq
We then use the algebraic equations to eliminate $F_{33}$, $F_{43}$, $F_{42}$, $F_{52}$ and $F_{51}$.
This leaves $F_{11}$, $F_{21}$, $F_{22}$, $F_{31}$, $F_{32}$ and $F_{41}$ which are determined from differential equations.
One obtains for the differential equation for the master integrals with the notation as in eq.~(\ref{A_f_notation})
for the first few terms of the $\qbar$-expansion
\bq
 f_{0,2} & = & 1,
 \nonumber \\
 f_{0,3} & = & 
 1 - 3 \qbar + 87 \qbar^2 - 2523 \qbar^3 + 74247 \qbar^4 - 2248278 \qbar^5 +69083151 \qbar^6+ \mathcal{O}(\qbar^7),
 \nonumber \\
 f_{0,4} & = &
 1 - 4 \qbar + 124 \qbar^2 - 3892 \qbar^3 + 123564 \qbar^4 - 3985904 \qbar^5 + 129468364 \qbar^6+ \mathcal{O}(\qbar^7),
 \nonumber \\
 f_{2,1} & = &
 -12 \qbar + 72 \qbar^2 - 1992 \qbar^3 + 45792 \qbar^4 - 1212912 \qbar^5 + 33130548 \qbar^6 + \mathcal{O}(\qbar^7),
 \nonumber \\
 f_{2,2} & = &
 -27 \qbar + 603 \qbar^2 - 19647 \qbar^3 + 634083 \qbar^4 - 20802702 \qbar^5 + 682840719 \qbar^6 + \mathcal{O}(\qbar^7),
 \nonumber \\
 f_{2,3} & = &
 \frac{21}{2} -87 \qbar + 2727 \qbar^2 - 95991 \qbar^3 + 3376767 \qbar^4 - 118926762 \qbar^5 + 4161308247 \qbar^6 
 \nonumber \\
 & &
 + \mathcal{O}(\qbar^7),
 \nonumber \\
 f_{4,1} & = &
 -12 \qbar + 612 \qbar^2 - 22692 \qbar^3 + 860292 \qbar^4 - 31443012 \qbar^5 + 1125105948 \qbar^6 + \mathcal{O}(\qbar^7),
 \nonumber \\
 f_{4,2} & = &
 -41 \qbar + 2921 \qbar^2 - 152933 \qbar^3 + 7213761 \qbar^4 - 314247466 \qbar^5 + 12916991381 \qbar^6 
 \nonumber \\
 & &
 + \mathcal{O}(\qbar^7),
 \nonumber \\
 f_{4,3} & = &
 -\frac{259}{4} -6 \qbar + 6096 \qbar^2 - 437658 \qbar^3 + 23412396 \qbar^4 - 1087900806 \qbar^5 + 46568896716 \qbar^6
 \nonumber \\
 & &
 + \mathcal{O}(\qbar^7),
 \nonumber \\
 f_{6,1} & = &
 -12 \qbar + 1692 \qbar^2 - 118812 \qbar^3 + 6760332 \qbar^4 - 338402412 \qbar^5 + 15469136748 \qbar^6 
 \nonumber \\
 & &
 + \mathcal{O}(\qbar^7),
 \nonumber \\
 f_{6,2} & = &
 \frac{735}{2} -723 \qbar + 10593 \qbar^2 - 129549 \qbar^3 + 5223333 \qbar^4 - 536169498 \qbar^5 + 39388876803 \qbar^6
 \nonumber \\
 & &
 + \mathcal{O}(\qbar^7),
 \nonumber \\
 f_{7,0} & = &
 5040\left(1 + 5 \qbar - 65 \qbar^2 + 725 \qbar^3 - 15825 \qbar^4 + 368530 \qbar^5 - 9202385 \qbar^6\right) + \mathcal{O}(\qbar^7),
 \nonumber \\
 f_{8,1} & = &
 114 \qbar - 13914 \qbar^2 + 772314 \qbar^3 - 31329954 \qbar^4 + 924096114 \qbar^5 - 13818576546 \qbar^6
 \nonumber \\
 & &
 + \mathcal{O}(\qbar^7),
 \nonumber \\
 f_{8,2} & = &
 -1624 - 6169 \qbar + 340489 \qbar^2 - 13341397 \qbar^3 + 463880769 \qbar^4 - 15021729194 \qbar^5
 \nonumber \\
 & &
 +478667081269 \qbar^6 +\mathcal{O}(\qbar^7),
 \nonumber \\
 f_{10,1} & = &
 6\left(147 + 425 \qbar - 255 \qbar^2 - 692305 \qbar^3 + 39335985 \qbar^4 - 1638625425 \qbar^5 
 \right. \nonumber \\
 & & \left.
 + 59749752435 \qbar^6\right) + \mathcal{O}(\qbar^7),
 \nonumber \\
 f_{12,1} & = &
 -144 \left(5 + 112 \qbar - 2197 \qbar^2 + 24217 \qbar^3 - 613167 \qbar^4 + 18230912 \qbar^5 - 521840698 \qbar^6\right) 
 \nonumber \\
 & & + \mathcal{O}(\qbar^7).
\eq
With the differential equation for the master integrals at hand, we obtain its solutions
with the boundary condition given by eq.~(\ref{boundary_value}).
The $\eps$-expansion of the master integral $M_1$ starts at order $\eps^6$:
\bq
 M_1 & = & 
 \eps^6 M_1^{(6,6)} + \eps^7 M_1^{(6,7)}
 + {\mathcal O}\left(\eps^8\right).
\eq
The first term in the $\eps$-expansion is given by
\bq
 M_1^{(6,6)}
 & = &
 1120 \zeta_3^2 -2016 \zeta_5 L_{\qbar} - 3360 \zeta_3 I\left(1,\Yinvariant_2,\Yinvariant_3\right) + I\left(1,\Yinvariant_2,\Yinvariant_3,\Yinvariant_2,1,f_{7,0}\right).
\eq
The first few terms of the $\qbar$-expansion of $M_1^{(6,6)}$ read
\bq
 M_1^{(6,6)}
 & = &
 1120 \zeta _3^2-560 \zeta _3 L_{\qbar}^3-2016 \zeta _5 L_{\qbar}+7 L_{\qbar}^6 
 + 210 \qbar (-32 \zeta _3+48 \zeta _3 L_{\qbar}-3 L_{\qbar}^4+8 L_{\qbar}^3  )
 \nonumber \\
 & &
 +\frac{105}{2}  {\qbar}^2 \left(208 \zeta_3-1392 \zeta_3 L_{\qbar}+87 L_{\qbar}^4-52 L_{\qbar}^3-180 L_{\qbar}^2-72 L_{\qbar}+192\right)
 \nonumber \\
 & &
 + {\mathcal O}\left(\qbar^3\right).
\eq
In fig.~\ref{fig_result_6_loops} we plot the results for $M_1^{(6,6)}$ and $M_1^{(6,7)}$ for $|x| > 49$.
We also plotted the results from the program $\verb|pySecDec|$ \cite{Borowka:2017idc}.
Again, we observe excellent agreement.


\section{Non-trivial relations}
\label{sect:relations}

In section~\ref{sect:picard_fuchs_operator} we considered the Picard--Fuchs operator $L^{(l,0)}$ and a Frobenius basis $\Frobeniusbasis^{(l)}_0, \dots, \Frobeniusbasis^{(l)}_{l-1}$.
We singled out the holomorphic solution $\Frobeniusbasis^{(l)}_0$ to normalise the master integral $M_1^{(l)}$,
the pair $(\Frobeniusbasis^{(l)}_1,\Frobeniusbasis^{(l)}_0)$ to define the mirror map and the ordered sequence
$(\Frobeniusbasis^{(l)}_0,\Frobeniusbasis^{(l)}_1,\dots,\Frobeniusbasis^{(l)}_{l-1})$ to define the $\Yinvariant$-invariants.

We know that at two loops we are not limited to this choice, we may choose any pair $(\tilde{\Frobeniusbasis}^{(2)}_1,\tilde{\Frobeniusbasis}^{(2)}_0)$ 
which generates the same lattice $\Lambda$.
As $(\tilde{\Frobeniusbasis}^{(2)}_1,\tilde{\Frobeniusbasis}^{(2)}_0)$ and $(\Frobeniusbasis^{(l)}_1,\Frobeniusbasis^{(l)}_0)$
generate the same lattice, they are related by a modular transformation
\bq
 \left( \begin{array}{c} \tilde{\Frobeniusbasis}^{(2)}_1 \\ \tilde{\Frobeniusbasis}^{(2)}_0 \\ \end{array} \right)
 & = & 
 \left( \begin{array}{cc} a & b \\ c & d \\ \end{array} \right)
 \left( \begin{array}{c} \Frobeniusbasis^{(2)}_1 \\ \Frobeniusbasis^{(2)}_0 \\ \end{array} \right),
 \;\;\;\;\;\;
 \left( \begin{array}{cc} a & b \\ c & d \\ \end{array} \right)
 \; \in \; \mathrm{SL}_2\left({\mathbb Z}\right)
\eq
and one finds that any choice leads to an $\eps$-factorised differential equation, with the entries of the matrix $A$ given by modular forms 
times the prefactor $\eps$ \cite{Weinzierl:2020fyx}.

We may now ask: Does this freedom of choice generalise to higher loops?
The answer is in general no. 
The conditions that terms of order $\eps^j$ with $j<1$ are absent (e.g. the conditions given in eq.~(\ref{cond_poles})) 
depend on properties, which are fulfilled for the choice we made in section~\ref{sect:picard_fuchs_operator}
(and possibly other choices), but not for arbitrary choices.
There are constraints.
We illustrate this with the simplest example, namely the function by which we normalise the master integral $M_1^{(l)}$.
Let us start from
\bq 
 \tilde{M}_1^{(l)} & = & \frac{\eps^l}{\psi} I_{1 \dots 1 1},
\eq
where a priori we treat $\psi$ as an arbitrary function of $y$.
From the condition that the pole of order $(l-4)$ of $A_{l 1}$ is absent
we learn that 
\bq
\label{homogeneous_Picard_Fuchs_eq_psi}
 L^{(l,0)} \psi & = & 0,
\eq
e.g. $\psi$ must be a solution of the homogeneous Picard--Fuchs equation.
This is expected.
The Picard--Fuchs equation is a linear differential equation of order $l$.
Eq.~(\ref{homogeneous_Picard_Fuchs_eq_psi}) implies that we may write $\psi$ as a linear combination with constant coefficients of the Frobenius basis
\bq
\label{linear_combination}
 \psi
 & = &
 \sum\limits_{j=0}^{l-1}
 c_j \Frobeniusbasis^{(l)}_j.
\eq
Is any such linear combination allowed? The answer is no. At odd loops and for $l \ge 3$ we find that $\psi$ must satisfy 
in addition to eq.~(\ref{homogeneous_Picard_Fuchs_eq_psi}) a non-linear differential equation of order $(l-1)$.
For three, five and seven loops these constraints read
 \begin{align}
    \label{eq:NLconstraints}
        \text{3L}:&\,\,\frac{1}{\psi} \frac{d^{2} \psi}{d y^{2}}
                       -\frac{1}{2}\left(\frac{1}{\psi} \frac{d \psi}{d y}\right)^{2}
                       +\frac{1}{2}\left(\frac{4}{1+4y}+\frac{16}{1+16y}\right) \frac{1}{\psi} \frac{d \psi}{d y}
                       +\frac{1+8y}{2 y^2(1+4y)(1+16y)}=0,\notag\,\\
        \text{5L}:&\,\,\frac{1}{\psi} \frac{d^{4} \psi}{d y^{4}}
                       -\frac{1}{\psi}\frac{d^{3} \psi}{d y^{3}}\left[\frac{1}{\psi}\frac{d \psi}{d y}
                                                                     -\frac{3}{2}\left(\frac{2}{y}+\frac{4}{1+4y}+\frac{16}{1+16y}+\frac{36}{1+36y}\right)\right]\notag\\
        &+\frac{1}{2}\frac{1}{\psi} \frac{d^{2} \psi}{d y^{2}}\left[\frac{1}{\psi} \frac{d^{2} \psi}{d y^{2}}
                                                                   -\frac{1}{\psi} \frac{d \psi}{d y}\left(\frac{2}{y}+\frac{4}{1+4y}
                                                                                                           +\frac{16}{1+16y}+\frac{36}{1+36y}\right)\right.\notag\\
        &\left.+\frac{4 \left(13824 y^3+2862 y^2+112 y+1\right)}{y^2 (1 + 4 y) (1 + 16 y) (1 + 36 y)}\right]
          +\frac{1}{\psi} \frac{d \psi}{d y}\left[-\frac{1}{\psi} \frac{d \psi}{d y}\frac{1 + 56 y + 1020 y^2 + 4608 y^3}{2 y^2 (1 + 4 y) (1 + 16 y) (1 + 36 y)}\right.\notag\\
        &\left.+\frac{4 (1 + 9 y) (7 + 192 y)}{y^2 (1 + 4 y) (1 + 16 y) (1 + 36 y)}\right]
         +\frac{1+12y}{2 y^4 (1 + 4 y) (1 + 16 y) (1 + 36 y)}=0,\,\notag\\
        \text{7L}:&\,\,\frac{1}{\psi} \frac{d^{6} \psi}{d y^{6}}
                       -\frac{1}{\psi} \frac{d^{5} \psi}{d y^{5}}\left[\frac{1}{\psi} \frac{d \psi}{d y}-\frac{5}{2}\left(\frac{4}{y}+\frac{4}{1+4y}
                                                                      +\frac{16}{1+16y}+\frac{36}{1+36y}+\frac{64}{1+64y}\right)\right]\,\notag\\
        &+\frac{1}{\psi} \frac{d^{4} \psi}{d y^{4}}\left[\frac{1}{\psi} \frac{d^{2} \psi}{d y^{2}}
                                                        -\frac{3}{2}\frac{1}{\psi} \frac{d \psi}{d y}\left(\frac{4}{y}+\frac{4}{1+4y}+\frac{16}{1+16y}
                                                                                                           +\frac{36}{1+36y}+\frac{64}{1+64y}\right)\right.\,\notag\\
        &\left.+\frac{2 (13 + 2628 y + 144150 y^2 + 2432512 y^3 + 9142272 y^4)}{y^2 (1 + 4 y) (1 + 16 y) (1 + 36 y) (1 + 64 y)}\right]
         +\frac{1}{\psi} \frac{d^{3} \psi}{d y^{3}}\left[-\frac{1}{2}\frac{1}{\psi} \frac{d^{3} \psi}{d y^{3}}\right.\,\notag\\
        &\left.+\frac{1}{2}\frac{1}{\psi} \frac{d^2\psi}{d y^2}\left(\frac{4}{y}+\frac{4}{1+4y}+\frac{16}{1+16y}+\frac{36}{1+36y}+\frac{64}{1+64y}\right)\right.\,\notag\\
        &\left.-\frac{1}{\psi} \frac{d\psi}{d y} \frac{4 (2 + 414 y + 22935 y^2 + 389696 y^3 + 1474560 y^4)}{y^2 (1 + 4 y) (1 + 16 y) (1 + 36 y) (1 + 64 y)}\right.\,\notag\\
        &\left.+\frac{6 (3 + 942 y + 69900 y^2 + 1489280 y^3 + 6782976 y^4)}{y^3 (1 + 4 y) (1 + 16 y) (1 + 36 y) (1 + 64 y)}\right]\,\notag\\
        &+\frac{1}{\psi} \frac{d^{2} \psi}{d y^{2}}\left[\frac{1}{\psi} \frac{d^{2} \psi}{d y^{2}}
                                                        \frac{1 + 228 y + 13110 y^2 + 228352 y^3 + 884736 y^4}{y^2 (1 + 4 y) (1 + 16 y) (1 + 36 y) (1 + 64 y)}\right.\,\notag\\
        &\left.-\frac{1}{\psi} \frac{d \psi}{d y} \frac{2 (1 + 342 y + 26220 y^2 + 570880 y^3 + 2654208 y^4)}{y^3 (1 + 4 y) (1 + 16 y) (1 + 36 y) (1 + 64 y)}\right.\,\notag\\
        &\left.+\frac{3 (1 + 492 y + 55544 y^2 + 1616896 y^3 + 9437184 y^4)}{y^4 (1 + 4 y) (1 + 16 y) (1 + 36 y) (1 + 64 y)}\right]\,\notag\\
        &+\frac{1}{\psi} \frac{d \psi}{d y}
                         \left[-\frac{1}{\psi} \frac{d \psi}{d y}\frac{1 + 108 y + 9312 y^2 + 283648 y^3 + 1769472 y^4}{2 y^4 (1 + 4 y) (1 + 16 y) (1 + 36 y) (1 + 64 y)}\right.\,\notag\\
        &\left.+\frac{6 (9 + 1552 y + 70912 y^2 + 589824 y^3)}{y^4 (1 + 4 y) (1 + 16 y) (1 + 36 y) (1 + 64 y)}\right]
        +\frac{1}{2 y^6 (1 + 4 y) (1 + 36 y) (1 + 64 y)} = 0.
 \end{align}
The choice $\psi=\Frobeniusbasis^{(l)}_0$ satisfies this constraint, but not every linear combination of the form as in eq.~(\ref{linear_combination}) does.

There is an interpretation of the constraint in the three-loop case:
At three loops we know that $L^{(3,0)}$ is a symmetric square, e.g. there exists a linear second-order
differential operator $\tilde{L}^{(2,0)}$ with solutions $\tilde{\psi}^{(2)}_0$ and $\tilde{\psi}^{(2)}_1$
such that
\bq
 \left\{ \left(\tilde{\psi}^{(2)}_0\right)^2, \tilde{\psi}^{(2)}_0 \tilde{\psi}^{(2)}_1, \left(\tilde{\psi}^{(2)}_1\right)^2 \right\}
\eq
span the solution space of $L^{(3,0)}$.
The constraint at three loops implies that $\psi$ has to be a perfect square \cite{Pogel:2022yat}, i.e. of the form
\bq
 \psi & = & \left(c_0 \tilde{\psi}^{(2)}_0 + c_1 \tilde{\psi}^{(2)}_1 \right)^2.
\eq
It is easily verified by direct computation that $\psi=\Frobeniusbasis^{(l)}_0$ satisfies the constraint in eq.~(\ref{eq:NLconstraints}).
Alternatively, this can be shown with the help of the quadratic relations satisfied by the Frobenius basis\footnote{We thank the anonymous referee for pointing this out.}.
Following \cite{Bonisch:2021yfw} we 
we define the $(l \times l)$ Wronskian matrix
\bq
 W & = &
 \left( \begin{array}{cccc}
  \psi_0 & \psi_1 & \dots & \psi_{l-1} \\
  \partial_y \psi_0 & \partial_y \psi_1 & \dots & \partial_y \psi_{l-1} \\
  \vdots & \vdots & & \vdots \\
  \partial_y^{l-1} \psi_0 & \partial_y^{l-1} \psi_1 & \dots & \partial_y^{l-1} \psi_{l-1} \\
 \end{array} \right)
\eq
and the $(l \times l)$ intersection matrix
\bq
 \Sigma
 & = &
 \left( \begin{array}{cccc}
  & & & 1 \\
  & & -1 & \\
  & 1 & & \\
  \iddots & & & \\
 \end{array} \right).
\eq
The quadratic relations satisfied by the Frobenius basis read \cite{Bonisch:2021yfw}
\bq
\label{quadratic_relations}
 W \Sigma W^T & = & Z,
\eq
where the entries of the $(l \times l)$ matrix $Z$ are rational functions of $y$.
Concretely, the entries are given as follows: We label the rows and columns of $Z$ from $0$ to $(l-1)$.
The entries in row zero are given by
\bq
 Z_{0 j} & = &
 \left\{ \begin{array}{ll}
  0, & j < l-1, \\
  \frac{1}{\left(2\pi i\right)^{l-1}} \alpha, & j = l-1,
 \end{array} \right.
\eq
with $\alpha$ given by eq.~(\ref{def_alpha_banana}).
The subsequent rows are then recursively constructed as follows:
\bq
 Z_{i j} & = &
 \left\{ \begin{array}{ll}
  \partial_y Z_{(i-1) j} - Z_{(i-1)(j+1)}, & j < l-1, \\
  \partial_y Z_{(i-1) j} + \sum\limits_{k=0}^{l-1} r^{(l,0)}_k Z_{(i-1)k}, & j = l-1,
 \end{array} \right.
\eq
where the $r^{(l,0)}_k$'s are the coefficients of the Picard-Fuchs operator $L^{(l,0)}$ in eq.~(\ref{def_Picard_Fuchs_eps_0}).
$\Sigma$ and $Z$ are symmetric matrices if $l$ is odd and skew-symmetric matrices if $l$ is even.
Let us now consider the case where $l$ is odd.
Eliminating from the set of equations given by eq.~(\ref{quadratic_relations}) the non-holomorphic solutions $\psi_1, \dots, \psi_{l-1}$
leads to eq.~(\ref{eq:NLconstraints}).


\section{Conclusions}
\label{sect:conclusions}

In this paper we presented a systematic method to transform 
the differential equation for the $l$-loop equal mass banana integral
into an $\varepsilon$-factorised form.
In particular this provides an example, that Feynman integrals related to Calabi--Yau $(l-1)$-folds
have a differential equation in $\eps$-factorised form.
With the known boundary value at a specific point this allows us for the banana integrals 
to obtain systematically the term of order $j$
in the expansion in the dimensional regularisation parameter $\varepsilon$ for any loop $l$.
The essential ingredient for our method is an ansatz for the master integrals, presented in section~\ref{sect_ansatz}.
We expect that with appropriate modifications this ansatz will be useful for Calabi--Yau Feynman integrals
beyond the family of banana integrals.

\subsection*{Acknowledgements}

We would like to thank Claude Duhr, Christoph Nega, Lorenzo Tancredi and Duco van Straten for useful discussions.
This work has been supported
by the Cluster of Excellence Precision Physics, Fundamental Interactions, and Structure of Matter, Universit\"at Mainz
and the Cluster of Excellence Origins, Technische Universität M\"unchen, both
funded by the German Research Foundation (DFG) within
the German Excellence Strategy (Project IDs EXC 2118-39083149 and EXC 2094–390783311).


\begin{appendix}


\section{The Picard--Fuchs operator from the Bessel representation}
\label{sect:bessel}

As a starting point we need the differential equation of the banana integrals
in some basis. This basis does not have to 
put the differential equation into an $\eps$-factorised form.
We could choose for example the derivative basis 
\bq
 I_{1 \dots 1 0},
 \;\;\;
 I_{1 \dots 1 1},
 \;\;\;
 \frac{d}{dy}I_{1 \dots 1 1},
 \;\;\;
 \dots,
 \;\;\;
 \frac{d^{l-1}}{dy^{l-1}}I_{1 \dots 1 1},
\eq
given in eq.~(\ref{def_derivative_basis}).
The system of these $(l+1)$ first order differential equations is equivalent to the Picard--Fuchs differential equation
of eq.~(\ref{full_Picard_Fuchs_dgl}).
In principle we may obtain the system of differential equations from standard
integration-by-parts reduction programs \cite{Smirnov:2008iw,Smirnov:2019qkx,Studerus:2009ye,vonManteuffel:2012np,Maierhoefer:2017hyi,Klappert:2020nbg}.
Such programs consider a larger graph, so that any scalar product involving any
loop momentum can be expressed as a linear combination of inverse propagators.
At $l$ loops this auxiliary graph has
\bq
 N_V & = & \frac{1}{2} l \left(l+3\right)
\eq
propagators.
This is the number of Baikov variables.
\begin{table}
\begin{center}
\begin{tabular}{|c|r|r|r|r|r|r|r|}
 \hline
 $l$   & $1$ & $2$ & $3$ & $4$ & $5$ & $6$ & $7$ \\
 \hline
 $N_V$ & $2$ & $5$ & $9$ & $14$ & $20$ & $27$ & $35$ \\
 \hline
\end{tabular}
\end{center}
\caption{
The number of propagators of the auxiliary graph at $l$ loops.
}
\label{table_number_Baikov_variables}
\end{table}
For low loop orders this number is shown in table~\ref{table_number_Baikov_variables}.
Standard integration-by-parts reduction programs are sufficient for $l \le 5$, but the large number of Baikov variables
becomes prohibitive at $l \gtrsim 6$.
We need a more efficient method.
This can be done based on an integral representation of banana integrals in terms of Bessel functions.
In the following we assume for simplicity $x<0$, the final result will be independent of this assumption.
The integral $I_{1 \dots 1 1}$ has the integral representation \cite{Berends:1993ee,Groote:2005ay}
\bq
\label{bessel_representation}
 I_{1 \dots 1 1}
 & = &
 e^{l \eps \gamma_E} 2^{l\left(1-\eps\right)} \left(-x\right)^{\frac{\eps}{2}}
 \int\limits_0^\infty dt \; t^{1+l\eps} J_{-\eps}\left(t\sqrt{-x}\right) \left[ K_{-\eps}\left(t\right) \right]^{l+1},
\eq
where $J_\nu(z)$ denotes the Bessel function of the first kind, and $K_\nu(z)$ denotes the modified Bessel function of the second kind.
We review a method which allows us to compute the Picard--Fuchs operator $L^{(l)}$ (including the $\eps$-dependent terms)
for higher loops \cite{Vanhove:2014wqa,Bonisch:2021yfw}.
This method is highly efficient and computes the Picard--Fuchs operator $L^{(15)}$ for the equal-mass banana integral with
$15$ loops in less than three seconds.
The starting point is a differential equation for $(K_{-\eps}(t))^{l+1}$:
We denote the Euler operator in the variable $t$ by $\theta_t = t \frac{\partial}{\partial t}$.
We first define recursively differential operators $B_{k l}$ through\footnote{The factor $(k-1)$ in front of the second term is missing in ref.~\cite{Bonisch:2021yfw}. It is a factor $k$ in their notation.}
\bq
 B_{0 l} \; = \;1, 
 \;\;\;
 B_{1 l} \; = \; \theta_t,
 \;\;\;
 B_{k l}
 \; = \;
 \theta_t B_{(k-1) l} - \left(k-1\right)\left(l-k+3\right) \left(t^2+\eps^2\right) B_{(k-2) l}.
\eq
The operator $B_{(l+2) l}$ annihilates $(K_{-\eps}(t))^{l+1}$ \cite{Bronstein:1997aaa,Borwein:2008aaa}:
\bq
 B_{(l+2) l} (K_{-\eps}(t))^{l+1} & = & 0.
\eq
In the next step we construct a differential operator $\tilde{B}_{l+2}$ in the variable $t$ such that
\bq
 \int\limits_0^\infty dt \; t^{1+l\eps} \left[ K_{-\eps}\left(t\right) \right]^{l+1} \tilde{B}_{l+2} J_{-\eps}\left(t\sqrt{-x}\right) 
 & = & 0.
\eq
Given the operator $B_{(l+2) l}$ this can be done with the help of integration-by-parts. 
The boundary terms vanish.
Given
\bq
 B_{(l+2) l}
 & = & 
 \sum\limits_{i=0}^{l+2} \sum\limits_{j=0}^{l+2-i} 
 b_{i j} t^j \theta_t^i
\eq
we obtain
\bq
 \tilde{B}_{l+2}
 & = &
 \sum\limits_{i=0}^{l+2} \sum\limits_{j=0}^{l+2-i} 
 \left(-1\right)^i b_{i j} t^j \left(\theta_t+j+2+\eps l\right)^i.
\eq
In the next step we convert from the operator $\tilde{B}_{l+2}$ in the variable $t$ to an operator
$\tilde{D}_{l+2}$ in the variable $x$ such that
\bq
 \tilde{D}_{l+2} \int\limits_0^\infty dt \; t^{1+l\eps} \left[ K_{-\eps}\left(t\right) \right]^{l+1} J_{-\eps}\left(t\sqrt{-x}\right) 
 & = & 0.
\eq
Here we use the relations
\bq
 \theta_t J_{-\eps}\left(t\sqrt{-x}\right)
 & = &
 2 \theta_x J_{-\eps}\left(t\sqrt{-x}\right),
 \nonumber \\
 t^2 J_{-\eps}\left(t\sqrt{-x}\right)
 & = &
 \frac{1}{x} \left( 4 \theta_x^2- \eps^2 \right) J_{-\eps}\left(t\sqrt{-x}\right).
\eq
The original integral in eq.~(\ref{bessel_representation}) has an additional factor $(-x)^{\frac{\eps}{2}}$ in front.
We define the differential operator $D_{l+2}$ such that
\bq
 D_{l+2} \left[\left(-x\right)^{\frac{\eps}{2}} \int\limits_0^\infty dt \; t^{1+l\eps} \left[ K_{-\eps}\left(t\right) \right]^{l+1} J_{-\eps}\left(t\sqrt{-x}\right) \right]
 & = & 0.
\eq
From the commutation relation
\bq
 \theta_x^n x^a & = & x^a \left(\theta_x+a\right)^n
\eq
it follows that $D_{l+2}$ is obtained from $\tilde{D}_{l+2}$ through 
the substitution $\theta_x \rightarrow \theta_x - \frac{\eps}{2}$.
We now have a differential operator $D_{l+2}$ of order $(l+2)$ in the variable $x$, which annihilates $I_{1 \dots 1 1}$:
\bq
 D_{l+2} I_{1 \dots 1 1} & = & 0.
\eq
The coefficient of the highest derivative $\theta_x^{l+2}$ is given by
\bq
 d_{l+2}
 & = &
 \left(-2\right)^{l+2} x^{\lfloor \frac{l+3}{2} \rfloor} \prod\limits_{a \in S^{(l)}} \left(x-a\right),
\eq
where $\lfloor x \rfloor$ denotes the largest integer $n$ with $n \le x$. 
The operator $D_{l+2}$ factorises as
\bq
 D_{l+2}
 & = &
 d_{l+2} L_{1,a} L_{1,b} L_x^{(l)},
\eq
where $L_{1,a}$ and $L_{1,b}$ are first-order differential operators and $L_x^{(l)}$ is the Picard--Fuchs operator in $x$-space.
The differential opeators $L_{1,a}$ and $L_{1,b}$ are given by
\bq
 L_{1,a}
 & = &
 \frac{d}{dx} 
 + \lfloor \frac{l+3}{2} \rfloor \frac{1}{x}
 + \sum\limits_{a \in S^{(l)}} \frac{1}{x-a}
 - \frac{\eps}{x} \theta\left(l \le 1\right),
 \nonumber \\
 L_{1,b}
 & = &
 \frac{d}{dx} 
 + \lfloor \frac{l+1}{2} \rfloor \frac{1}{x}
 + \sum\limits_{a \in S^{(l)}} \frac{1}{x-a}
 - \frac{\eps }{x} \theta\left(l > 1\right).
\eq
Here, $\theta(l \le 1)$ and $\theta(l > 1)$ denote Heaviside step functions.
Note that the distribution of the $\eps$-dependent terms differs for $l\le1$ and $l>1$.
Given $D_{l+2}$ and the known forms of $L_{1,a}$ and $L_{1,b}$, we obtain the Picard--Fuchs operator $L_x^{(l)}$ by
left-division with $L_{1,a}$ and $L_{1,b}$.
Note that left-division is significantly faster than factorisation of $D_{l+2}$.
Finally, a change of variables
\bq
 x \; = \; - \frac{1}{y},
 & &
 \frac{d}{dx} \; = \; y^2 \frac{d}{dy}
\eq
and division by $y^{2l}$
converts $L_x^{(l)}$ from $x$-space to the Picard--Fuchs operator $L^{(l)}$ in $y$-space.

\end{appendix}

\bibliography{/home/stefanw/notes/biblio}
\bibliographystyle{/home/stefanw/latex-style/h-physrev5}

\end{document}